\documentclass[9pt,journal]{IEEEtran}

\usepackage{amsmath,epsfig,stmaryrd}
\usepackage{color}
\usepackage{dsfont}

\usepackage{caption}
\usepackage{lipsum}
\usepackage{algorithmic}

 \usepackage{amssymb}
\usepackage{soul}
\usepackage{mathrsfs}

\usepackage{setspace}

\usepackage{mathtools}

\usepackage{amsmath,amsfonts,amssymb,rotating}

\mathcode`l="8000
\begingroup
\makeatletter
\lccode`\~=`\l
\DeclareMathSymbol{\lsb@l}{\mathalpha}{letters}{`l}
\lowercase{\gdef~{\ifnum\the\mathgroup=\m@ne \ell \else \lsb@l \fi}}%
\endgroup

\renewcommand{\[}{\left[}
\renewcommand{\]}{\right]}
\newcommand{\lb}{\llbracket}
\newcommand{\rb}{\rrbracket}

\def\P{{\mathbb P}}

\newfont{\bbb}{msbm10 scaled 500}

\newfont{\bb}{msbm10 scaled 1100}

\newcommand{\bv}{{\bf b}}

\newcommand{\iv}{{\bf i}}

\newcommand{\pv}{{\bf p}}

\newcommand{\xv}{{\bf x}}

\newcommand{\Xv}{{\bf X}}   

\newcommand{\Bc}{{\cal B}}

\newcommand{\Ec}{{\cal E}}
\newcommand{\Fc}{{\cal F}}

\newcommand{\Jc}{{\cal J}}

\newcommand{\Xc}{{\cal X}}

\newcommand{\js}{(j^*)}
\renewcommand{\j}{(j)}
\renewcommand{\k}{(k)}
\renewcommand{\l}{(l)}
\renewcommand{\L}{(L)}
\newcommand{\Ejs}{\Ec^*}

\newcommand{\Xl}{X_{(l)}}
\newcommand{\Xk}{X_{(k)}}
\newcommand{\Xjs}{X_{(j^*)}}
\newcommand{\Xj}{X_{(j)}}

\newcommand{\Jsetk}{\mathcal{J}_{(k)}}
\newcommand{\Jsetkd}{\mathcal{\widetilde{J}}_{(k)}}

\newcommand{\Mk}{M_{(k)}}
\newcommand{\M}{M}
\newcommand{\Rkjs}{R_{(k)|(j^*)}}
\newcommand{\RkEjs}{R_{(k)|\Ec^*}}
\newcommand{\Ffamk}{\mathcal{F}_{(k)}}
\newcommand{\sigA}{\pi}	
\newcommand{\Sk}{S_{(k)}}

\newcommand{\Pset}{\mathscr{P}}

\newcommand{\xkJxj}{_{\xv_{\k}^n,\xv_{\j}^n}}
\newcommand{\xkxj}{_{\xv_{\k}^n|\xv_{\j}^n}}
\newcommand{\xkxjs}{_{\xv_{\k}^n|\xv_{\js}^n}}

\newcommand{\SetET}{\boldsymbol{\hat{\theta}}^{(u_n)}}
\newcommand{\SetETR}{\boldsymbol{\widehat{\Theta}}^{(u_n)}}

\newcommand{\SetER}{\boldsymbol{\widehat{\Theta}}}
\newcommand{\An}{A_{n}(\mathbf{X}^{n},\Xv_{\j}^{n})}
\newcommand{\Bn}{B_{n}(\mathbf{X}^{n},\Xv_{\j}^{n})}
\newcommand{\Vn}{V_{n}(\mathbf{X}^{n},\Xv_{\j}^{n})}

\newcommand{\Pt}{Q}
\newcommand{\pt}{q}

\newcommand{\Hs}{\overline{H}}
\newcommand{\plim}{\text{p}\text{ - }\underset{n\rightarrow \infty}{\lim\sup}~}
\newcommand{\prlim}{\text{p}\text{ -}\underset{n\rightarrow \infty}{\lim}~}

\newcommand{\sigx}{\sigma_k}
\newcommand{\sigy}{\sigma_j}
\newcommand{\cc}{\rho_{k,j}}
\newcommand{\Ckj}{\Sigma_{(k,j)}}
\newcommand{\sigkj}{\sigma_{k,j}}
\newcommand{\sigkjs}{\sigma_{k,j^{*}}}
\newcommand{\Un}{\mathbf{U}^n}
\newcommand{\un}{\mathbf{u}^n}

\newcommand{\Reqvec}{\mathbf{K}} 
\newcommand{\NReq}{K} 
\newcommand{\reqt}{(k_t)} 
\newcommand{\reqtprev}{(k_{t-1})} 

\newcommand{\G}{\mathcal{G}} 
\newcommand{\V}{\mathcal{V}} 
\newcommand{\Edge}{\mathcal{E}} 

\renewcommand{\arg}{{\hbox{\rm arg}}}

\newcommand{\<}{\left\langle}
\renewcommand{\>}{\right\rangle}

\def\<{\langle}
\def\>{\rangle}

\newtheorem{theorem}{Theorem}

\newtheorem{definition}[theorem]{Definition}

\newtheorem{lemma}[theorem]{Lemma}

\ifCLASSINFOpdf

\else

\fi

\begin{document}
\onecolumn

\title{
Transmission and Storage Rates for Sequential Massive Random Access}

\author{Elsa Dupraz$^{1}$, Thomas Maugey$^2$, Aline Roumy$^2$, Michel Kieffer$^3$ \\ ~ \\ 
\small $^1$ Telecom Bretagne; UMR CNRS 6285 Lab-STICC, \small $^2$ INRIA Rennes Bretagne-Atlantique, $^3$ L2S, UMR CNRS 8506; CentraleSupelec; Univ. Paris-Sud}

\maketitle

\IEEEpeerreviewmaketitle

%
%

%
\begin{abstract}
This paper introduces a new source coding paradigm called Sequential Massive Random Access (SMRA). In SMRA, a set of correlated sources is encoded once for all and stored on a server, and clients want to successively access to only a subset of the sources. Since the number of simultaneous clients can be huge, the server is only allowed to extract a bitstream from the stored data: no re-encoding can be performed before the transmission of the specific client's request. In this paper, we formally define the SMRA framework and  introduce both storage and transmission rates to characterize the performance of SMRA. We derive achievable transmission and storage rates for lossless source coding of i.i.d. and non i.i.d. sources, and transmission and storage rates-distortion regions for Gaussian sources. We also show two practical implementations of SMRA systems based on rate-compatible LDPC codes. Both  theoretical and  experimental results demonstrate that SMRA systems can reach the same transmission rates as in traditional point to point source coding schemes, while having a reasonable overhead in terms of storage rate. These results constitute a breakthrough for many recent data transmission applications in which different parts of the data are requested by the clients.
\end{abstract}

\begin{IEEEkeywords}
Massive Random Access, Free-viewpoint Television, Video Compression, Sensor Networks, Lossless and Lossy Source Coding.
\end{IEEEkeywords}

%
\section{Introduction}

According to  \cite{Hilbert_M_2011_science}, the amount of data available on the web is growing exponentially. 
In the huge databases of, \emph{e.g.}, videos or pictures stored in online servers, the data usually contains redundancies. 
Efficient coding algorithms have been developed since decades to exploit these redundancies in order to limit the tremendous increase of storage needs.
In applications such as free-viewpoint television (FTV) or when exploiting data collected by a network of sensors, the items of the database (the different views of a video or the signals captured by different sensors) show strong correlations with each other.
In such applications, all the database should be compressed jointly in order to exploit all the redundancies and hence to maximize the storage efficiency.

In emerging applications such as FTV, the data is massively and interactively accessed  by heterogeneous clients.
The explosion of the data volume makes unrealistic and overall not desirable for the client the reception of the whole dataset. 
For example, in FTV~\cite{Tanimoto_2012_ieee-spm_ftv_fvt} or in sensor networks, the client is only interested by a subset of the views or of the collected data, and this subset may differ among clients.
In such schemes, a large number of sources are jointly compressed, and different subsets of them must be transmitted to the clients, depending on their needs. 
In order for traditional coding schemes to be efficient both in terms of storage and transmission rates, the server would need to decode the whole stream and re-encode only the requested sources. 
However, in this context, re-encoding is not desirable due to the potentially large number of simultaneous clients' requests.
As a result, for this new context of data delivery, traditional coding schemes cannot minimize both the storage rate and the transmission rate without re-encoding.  
Hence, there is a need for the development of new performing coding algorithms that take into account the variety of client's requests.

While some applications of the described scenario already exist, they have never been formally studied, and more particularly their transmission and storage performance have never been derived. In this paper, we propose a formal definition and a theoretical analysis of this new data coding and delivery scheme where only part of the sources are requested by the clients.
We call this paradigm, the \emph{Sequential Massive Random Access (SMRA)}, where ``Random" refers to the uncertainty in the clients requests, ``Massive" describes the fact that the number of clients requesting data is high and makes the data re-encoding unfeasible, and ``Sequential" means that clients request sources one after the other and can keep them in memory in order to improve the decoding efficiency.

The key novel aspect of SMRA is to study together the storage and transmission issues, while in the current state-of-the-art of information theory, storage and transmission are studied separately.
One of the main difficulties of SMRA is to take into account the uncertainty of the successive client's requests, \emph{i.e.} the subset of sources that might be extracted from the stored coded bitstream is random. 
With this uncertainty, standard coding solutions either have to do re-encoding which leads to a high complexity at the server, or to consider all the possible sets of requests and store all the corresponding encoded bitstreams on a server with a high storage cost. 
%
 In other words, the problem that is posed by SMRA is the following. 
Given a set of sources, can we encode them once and only once at a coding rate that minimizes the storage cost, while expecting a rate-efficient extraction of a subset of these sources? 
Or, if a rate-efficient compression is targeted, how much data should be stored in the bitstream that is \emph{a priori} coded?

\subsection{Overview of the main contributions}
Three main contributions are presented in this paper:
\begin{enumerate}
\item \emph{SMRA definition and new coding scheme:} we formally define the SMRA problem  in Section~\ref{sec:MRA}. 
We first introduce the new context of SMRA systems, that is a set of sources encoded once for all and stored on a server, and clients asking for subset of sources one after the other. 
In this case, when a client sends a request to the server at a given time instant, it is assumed that its previous request is still in its memory.
A coding scheme for SMRA is then proposed, involving a two-step encoding strategy: \emph{offline encoding} for the storage of sources, and \emph{online extraction} that transmits to the client only the information he needs with respect to his request and without re-encoding. 
The problem of minimizing the storage rate (from offline encoding) and the transmission rate (from online extraction) of SMRA  is then formulated in the case when the one previously requested source is stored in the client's memory.

\item \emph{Theoretical resolution of SMRA problem:} in Section~\ref{sec:soa}, we first revisit state-of-the-art results that are related to the SMRA problem. They permit to give insights on the transmission and storage rates that can be achieved by our coding scheme. In Section~\ref{sec:RSbounds:iid}, we then provide a complete information-theoretical study of the transmission and storage rates that can be achieved by SMRA. We, in particular, derive new information-theoretic bounds on these rates for non i.i.d. sources  (Section \ref{sec:RSbounds:gen}), no knowledge of source statistics (Sections \ref{sec:iid_stat_unknown} and \ref{sec:RSbounds:gen}), lossy compression (Section \ref{sec:lossy}), and unbounded memory  (Section \ref{sec:RSboundsMRA:iid}) (not one, but all the previously requested source are stored in the client's memory). 
In all the considered cases, we show that the minimum achievable storage rate  is obtained by taking into account all the possible paths from previous request to current request and is given by the worst one, \emph{i.e.}, the one presenting the lowest correlation between the successively requested sources. In addition, we show that it is possible to construct the stored bitstream so as to achieve a transmission rate equal to the rate needed when re-encoding is possible. 
We also show that the lack of knowledge on source statistics does not increase the storage rate nor the transmission rate. 
%
\item \emph{Practical resolution of SMRA problem:} finally, we propose practical coding schemes that permit to reach these theoretical rates when particular source models are considered. In Section~\ref{sec:channelcoding}, we explain how incremental channel coding can be used to build computationally efficient algorithms. In Section~\ref{sec:exp}, we consider two particular source models and describe practical implementations of SMRA systems based on Low-Density Parity Check (LDPC)-staircase \cite{Roca12RFC6816} and Low-Density Parity Check Accumulator (LDPCA)~\cite{Varodayan06Eurasip} codes.

\end{enumerate}

\subsection{Related problems}
The SMRA framework defined in this paper shows connections with other problems that have been studied in the field of multi-source compression.
First, in the problem of data delivery with caching, a server makes available a database, and each of the clients involved in the network also stores a limited part of the content of the database.
Then, during the delivery phase, the clients request some data to the server, that has to send back the requested data through a single link shared by all the clients. 
When studying the caching problem, the objective is to minimize the delivery rate that is the amount of data transmitted from the server to the client, under constraints on the client's memory that is the amount of data stored by the clients.
For this problem, theoretical rate-memory regions were derived in~\cite{maddah14IT} for i.i.d. sources, and in~\cite{hassanzadeh16arxiv} for correlated sources, where the correlation is between sources.
In the SMRA framework, the transmission rate has the same definition as for caching, but the storage rate corresponds to the amount of data stored by the server, while in caching, it corresponds to data stored by the clients.
As another difference in the system definition, the caching framework considers only one single common transmission link from the server to all clients, while SMRA assumes direct links from the server to each of the clients. 
The theoretical approaches also differ, since in the caching formulation, the data is assumed to be already in a compressed form and each item of the database is compressed individually and separately from the others.
The main issue of caching is hence to deliver the subset of data that corresponds to the clients' requests with only one common link. 
Hence the source coding aspects are not discussed in the caching framework while they are the central point of our analysis.
In the same way, distributed storage presented in~\cite{dimakis10IT} is interested in duplication and data repair in case of server failure, but does not consider the source coding aspects of data storage.

From a source coding perspective, several works show connections with the SMRA framework.
First, in source coding with multiple descriptions~\cite{Gamal82IT,diggavi04ITW}, the source is compressed into different packets such that one packet is sufficient to reconstruct the source, but receiving more packets improves the quality of the reconstruction.
The multiple description problem is suboptimal in the sense that the rate needed to reach a given distortion $D$ when several packets are received is higher than the standard rate-distortion function of the source.
The suboptimality comes from the constraint that each individual packet must be sufficient to reconstruct the source with a sufficient quality.
On the opposite, SMRA corresponds to a problem with one single description from which we want to extract some data that corresponds to the client's request.

The information-theoretic problems that are the closest to SMRA are source coding with multiple decoders~\cite{sgarro77IT}, and source coding with unknown statistics between the source and the side information~\cite{Draper04}.
In the SMRA scheme, the side information is the previous request which is assumed to be in the user's memory. 
Standard problems of source coding with side information either study the storage rate~\cite{sgarro77IT} or the transmission rate~\cite{Draper04}, but none of them considers the joint design of these two quantities. 
The analysis of these results however gives insights on the transmission and storage rates that can be jointly achieved by SMRA, and this is why we restate them in Section~\ref{sec:MRA}.

\subsection{Notations}
In this paper, a random source $X$ is denoted using uppercase; the source $X$ generates a sequence of random variables denoted $X_i$  using uppercase and index $i$ while its realization $x_i$ is denoted using lowercase; a random vector $\Xv$ is denoted using boldface uppercase and its realization $\xv$ is denoted using boldface lowercase. An $n$-length vector $\Xv^n=(X_1,...,X_n)$ containing elements $X_1$ to $X_n$ is denoted using superscript $n$.
The alphabet $\Xc$ of a random variable  is denoted with calligraphic letter, and with the same letter as the random variable. $|\Xc|$ denotes the cardinality of the set $\Xc$. 
If $\Bc$ is a finite source alphabet, then $\Bc^{+}$ is the set of all finite strings from $\Bc$.
 $\forall \mathbf{b} \in \Bc^{+}$, $|\bv|$ stands for the length of $\bv$.

In the case of multiple sources, the set $\Jc$ of source indexes is denoted using calligraphic letter and the set of sources with indexes in $\Jc$ is denoted $X_{\Jc}$. The $k^{th}$ source is then identified with an index inside brackets i.e. $_ {(k)}$.
$\llbracket 1,N \rrbracket$ stands for the set of integers between $1$ and $N$.
%
\section{Sequential Massive random access problem}
\label{sec:MRA}
In this section, we define the general \emph{Sequential Massive Random Access} (SMRA) problem. We then describe Free Viewpoint Television (FTV) as a particular instance of SMRA.

\subsection{Graph-based definition of SMRA framework} \label{sec:MRAdefinition}
\begin{figure}[h]\begin{center}
\includegraphics[width=0.7\linewidth]{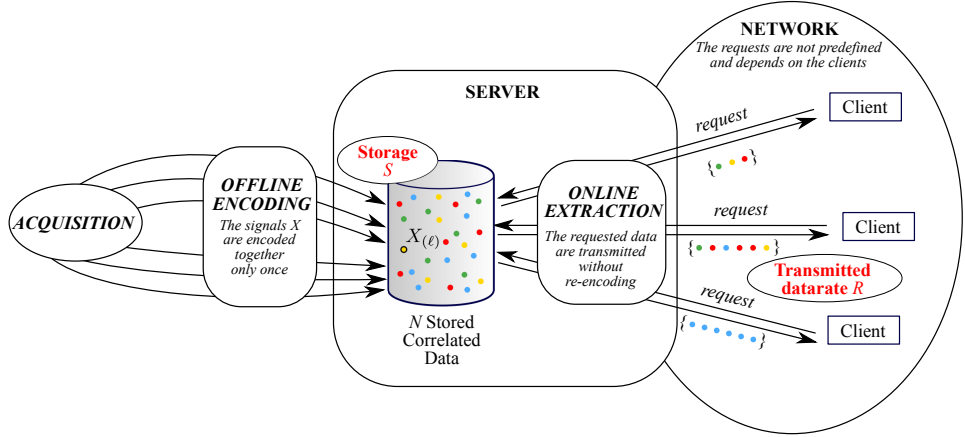}
\end{center}
\caption{SMRA framework}\label{fig:MRA_framework}\end{figure}

The SMRA framework is illustrated in Figure~\ref{fig:MRA_framework}.
Let $\{\Xl\}_{1\le l \leq L}$ be a set of $L$ sources $\Xl = (\Xv_{(l)}^n)_{n\rightarrow \infty}$. In the following, the sources can be \emph{general} or \emph{i.i.d.} as defined below.

\begin{definition}[Discrete general sources]
Let $\Xc$ be a discrete alphabet. 
The set $ \{\Xl\}_{1\le l \le L}$ is said to be a {\em set of $L$ discrete general sources} if the $L$  sources generate random vectors of size $n\ge 1$ according to the joint probability distribution $ P(\xv_{(1)}^n, \cdots, \xv_{(L)}^n)$, where $\xv_{(1)}^n, \cdots, \xv_{(L)}^n \in \mathcal{X}^n $ ,$\forall n\ge1$.
\label{def:generalSources}
\end{definition}

\begin{definition}[Discrete i.i.d. sources]
Let $\Xc$ be a discrete alphabet. 
The set $\{\Xl\}_{1\le l \le L}$ of general sources  is said to be a {\em set of $L$ discrete i.i.d. sources} if the joint probability distribution can be factorized as:
\begin{equation}\label{eq:factoriid}
\forall n\ge1, \ \ P(\xv^n_{(1)},...,\xv^n_{\L}) = \prod_{i=1}^n P(x_{(1),i},...,x_{\L,i})
\end{equation}
where $P(x_{(1),i},...,x_{\L,i})$ does not depend on $i$, and $x_{(\ell),i} \in \mathcal{X}$, $\forall i \in \llbracket 1,N \rrbracket, \ell \in \llbracket 1,L \rrbracket$.
\label{def:iidSources}
\end{definition}

As shown in Fig.~\ref{fig:MRA_framework}, the $n$ symbols of each source are acquired, encoded, and stored as a bitstream on a \emph{server}. These three operations are done \emph{a priori} (\emph{i.e.,} offline), which means that they are done once, without knowing which of the sources will be requested by clients. 
At the other side of the network, the clients want to sequentially access some of the sources, and thus send consecutive requests to the server. 

Consider a  client sequentially asking for $\NReq \leq L$ sources. 
The request takes the form of a vector $\Reqvec$, such that $\Reqvec = \[k_0, k_1, \ldots, k_t, \ldots, k_K \]$, where $k_t  \in \{1,\cdots, L\}$ for $t\geq 1$ and $k_0 = 0$. $X_{(0)}$  is, by convention for initialization, a source with zero entropy. The vector $\Reqvec$ indicates that the source of index $k_1$ is requested first, and that the source of index $k_t$ is requested just after the source of index $k_{t-1}$ for all $1\leq t \leq \NReq$. 
Since the successive indices $k_t$ design a path through the source set, we call the request vector $\Reqvec$ the \emph{client navigation}.
The index $t$ in  $k_t$ is called the \emph{navigation instant}. A request to the source of index $k_t$ corresponds to a request to all the $n$ symbols of $\xv_{\reqt}^n$. 
 
\begin{figure}[h]\begin{center}
\includegraphics[width=0.4\linewidth]{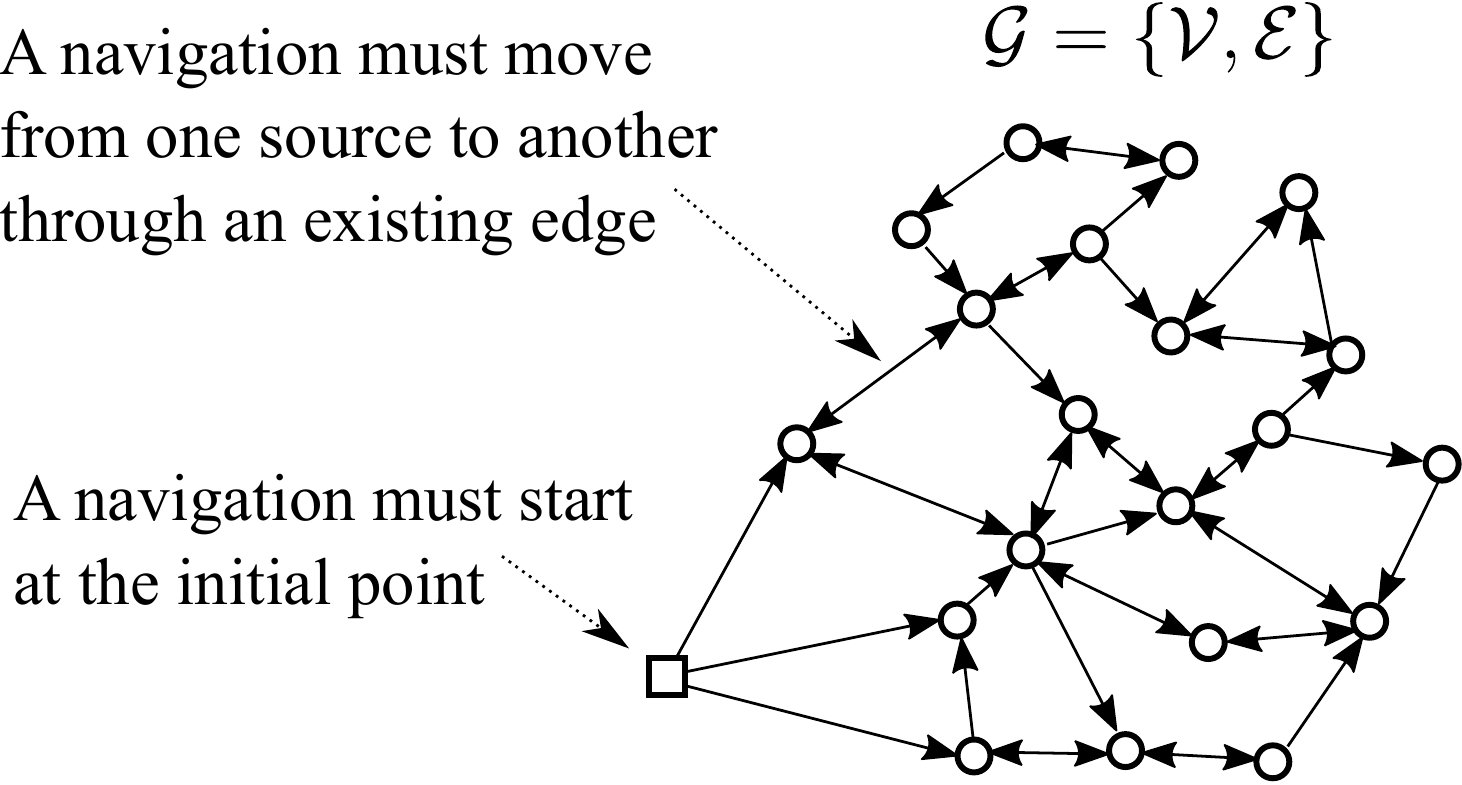}
\end{center}
\caption{The possible navigations in the context of SMRA can be represented with a graph $\G = \{\V,\Edge\}$. Each vertex is labeled by an integer belonging to $0\leq l \leq L$ and represents a source. An edge in $\Edge$ from $\ell$ to $\ell'$ indicates that a client can have access to source $X_{(\ell')}$, once he has accessed source $X_{(\ell)}$.
The graph root represents the source $X_{(0)}$ of zero entropy introduced for initialization and is represented by a square. The $L$ other sources are represented by circles.
}\label{fig:graphMRA}\end{figure}
 
In general, the client's navigation is constrained by a structure inherent to the set of sources $\{\Xl\}_{1\le l \leq L}$ and that depends on the application (\emph{e.g.} FTV). 
This structure is given \emph{a priori} and can be described by an oriented rooted graph $\G = \{\V,\Edge\}$, in which  the set of vertices $\V$ represents the $L+1$ sources (including $X_{(0)}$ corresponding to the root), \emph{i.e.,} $\V = \{0,1,\ldots,L\}$. 
There is an edge from vertex $l$ to vertex $l'$ if a client is allowed to request the source of index $l'$ just after having requested the source of index $l$. 
This is summarized in Figure~\ref{fig:graphMRA}, where the root is depicted as a square.

\begin{figure}[h]\begin{center}
\includegraphics[width=0.6\linewidth]{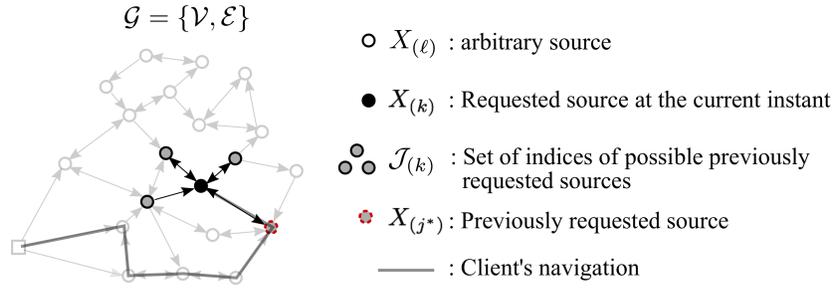}
\end{center}
\caption{SMRA notations where the root (vertex $0$) is denoted by a square.}\label{fig:notations}\end{figure}

In the SMRA framework, we assume that the client  still has in memory the previously requested source $X_{\reqtprev}$ when he receives his current requested source $X_{\reqt}$\footnote{For the sake of clarity, we assume here that the client has only one source in memory: the one requested at the previous instant. The more general case where several requested and memorized sources is studied in Section~\ref{sec:RSboundsMRA:iid}.}. 
Therefore, using the graph structure, the encoder knows that if a client requests a source of index $k_t$ with $t > 1$, its previous request $k_{t-1}$ is necessarily a neighbor of vertex $k_t$ in the graph $\G$. In other words, the processing of source $X_{\reqt}$ depends only on the sources in its neighborhood.
This enables us, in the following, to drop the navigation instant $t$ by considering only two navigation temporalities: the current navigation instant and the previous one.
More formally, we assume that, at the \emph{current navigation instant}, a client wants to access the source of index $k$. $\Xk$ is thus called the \emph{requested source}. At the \emph{previous instant}, the indices of possible \emph{previously requested sources} are the neighbors of vertex $k$ in the graph $\G$ and they are gathered in a set denoted by $\Jsetk$. Moreover, the actual previously requested source is denoted by $\Xjs$ with $j^*\in\Jsetk$.

 In the next section, we give an interpretation of these notations for one practical example of SMRA system:  FTV.

\subsection{Free-viewpoint television: an example of SMRA system}

\begin{figure}[h]\begin{center}
\includegraphics[width=0.6\linewidth]{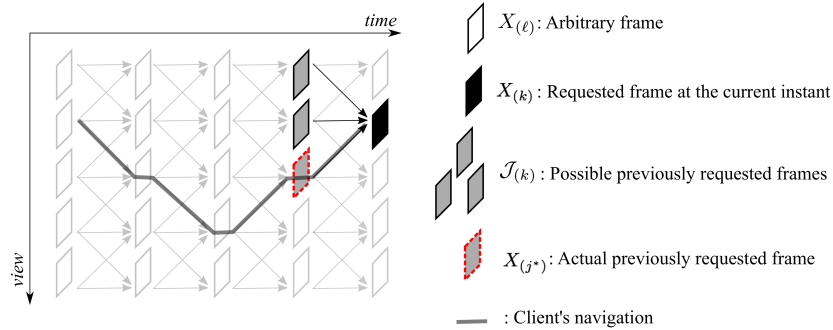}
\end{center}
\caption{The free viewpoint navigation is a particular example of SMRA.}\label{fig:ftv_ra}\end{figure}

As stated in introduction, FTV \cite{Tanimoto_2012_ieee-spm_ftv_fvt} is a particular instance of SMRA. In FTV, as illustrated in Figure~\ref{fig:ftv_ra}, a client can navigate between different views of a video, which means that at each instant, he picks one frame $\Xk$ (\emph{i.e.}, one image belonging to one view) among all the possible ones $\{\Xl\}_{1\le l \leq L}$.
In this model, each frame $\Xl$ is seen as a source; each $\Xv_{(l)}^n$ being an image (\emph{i.e.,} a set of pixels). 
Considering a free viewpoint acquisition system of $C$ cameras acquiring each $M$ successive frames, one obtains a set of $L=C\times M$ sources.

According to the notations of Section~\ref{sec:MRAdefinition}, the current requested frame is $\Xk$ (in black in Figure~\ref{fig:ftv_ra}) and the client has already requested one frame $\Xjs$, with $j^* \in \Jsetk$ (in grey in Figure~\ref{fig:ftv_ra}).
The set $\Jsetk$ depends on the navigation degree of freedom that is given to the client, which defines the graph $\G$ represented by the arrows in Figure~\ref{fig:ftv_ra}.
For instance, as an extreme case, the client may be allowed to request any frame from any time of any camera. In this case, the set $\Jsetk$ contains all the frame indices $\{0, 1,\ldots, L\}$ and the graph $\G$ is fully connected. 
Alternatively, as in most standard navigation scenarios, the client may only be able to move to one of the next views at each time instant \cite{Maugey_T_2013_tmm_int_mvsncnd}. In the latter case, the set $\Jsetk$ consists of three frame indices, as depicted in Figure~\ref{fig:ftv_ra}. 


\subsection{Coding scheme definition for SMRA}\label{sec:problem}

\begin{figure}[h]\begin{center}
\includegraphics[width=0.6\linewidth]{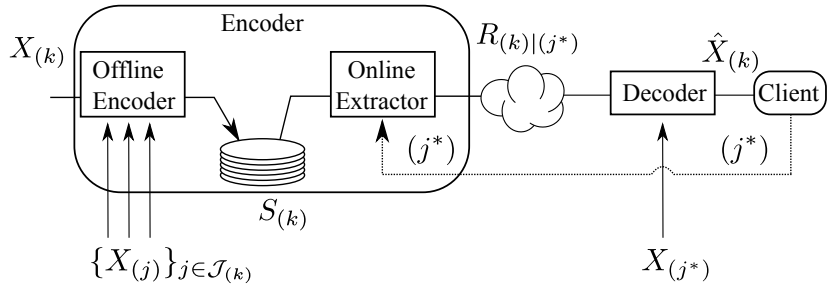}
\end{center}
\caption{SMRA: focus on the coding of one source $\Xk$}\label{fig:MRA_problem}\end{figure}

Consider the SMRA problem depicted in Figure~\ref{fig:MRA_problem}, where only one source $\Xk$ is to be recovered losslessly. 
During the first phase, the source $\Xk$ is encoded  (compressed) by the \emph{offline encoder} into an index sequence at storage rate $\Sk$ bits per source symbol under the assumptions that (i)  all the realizations of the sources in $\Jsetk$ are known, (ii)  one source from  $\Jsetk$ will be available at the client's decoder. 
The index sequence is communicated over a noiseless link  to a server. 
Then, the second phase starts when the client requests the source of index $k$ and specifies the index $j^* \in \Jsetk$ of the source already available at its decoder. Upon the client's request, the server extracts an index subsequence at rate $\Rkjs$ (\emph{online extractor}) and sends it over a noiseless link to the \emph{decoder} which finds an estimate of the requested source.  We now define formally the SMRA code and the achievable rates for this code.

\begin{definition}[SMRA code] 
\label{def:SMRAcode}
A $( (2^{n\Sk}, ( 2^{n \Rkjs} )_{j^* \in \Jsetk } )_{1 \le k \le L},n)$ \emph{SMRA code} for the set of discrete general sources $\{ \Xk \}_{1\leq k \leq L}$  consists, for each source $\Xk$, $1 \leq k \leq L$, of
\begin{itemize}
\item an \emph{offline encoder} $h^{\text{off}}_{\k}$ that assigns a sequence of $\Mk=|\Jsetk|$ indices 
to each possible set of vectors $\left(\xv^n_{\k}, ( \xv^n_{\j}  )_{j \in \Jsetk }\right) \in \Xc^n \times \Xc^{n\times\Mk}$ 
\begin{subequations}\label{eq:enc-off}
\begin{align}
h^{\text{off}}_{\k}: \Xc^n \times \Xc^{n \times \Mk} &\to  \prod_{m=1}^{\Mk} \{ 1, \ldots, 2^{nr_{k,m}} \} \\
\xv^n_{\k}, ( \xv^n_{\j}  )_{j \in \Jsetk } & \mapsto ( i_1, \ldots , i_{\Mk} ) 
\end{align}
\end{subequations} 
where $\Sk = r_{k,1} + r_{k,2} + \ldots + r_{k,\Mk}  $.
\item a set of $\Mk$ \emph{online extractors} $h^{\text{on}}_{{(k)|(j^*)}}$, $j^* \in \Jsetk$, that extract a subsequence of indices from the sequence of indices $( i_1, \ldots , i_{\Mk} )$
\begin{subequations}\label{eq:enc-on}
\begin{align}
h^{\text{on}}_{{(k)|(j^*)}}: \prod_{m=1}^{\Mk} \{ 1, \ldots, 2^{nr_{k,m}}\}&\to  \prod_{m \in \mathcal{I}_{(k)|(j^*)} } \{ 1, \ldots, 2^{nr_{k,m}} \} \\
 ( i_1, \ldots , i_{\Mk} ) & \mapsto ( i_m )_{m \in \mathcal{I}_{(k)|(j^*)}}
\end{align}
\end{subequations} 
where $\mathcal{I}_{(k)|(j^*)} \subseteq \{1, \cdots, \Mk\}$, and $\Rkjs = \sum_{m \in \mathcal{I}_{(k)|(j^*)}} r_{k,m}  \le \Sk$
\item a set of $\Mk$ \emph{decoders} $g_{{(k)|(j^*)}}$, $j^* \in \Jsetk$, that, given the source realization   $\xv^n_{\js} $, assigns an estimate $\hat \xv^n_{\k|\js}$ to each received subsequence of indices
\begin{subequations}\label{eq:dec}
\begin{align}
g_{{(k)|(j^*)}}:    \prod_{m \in \mathcal{I}_{(k)|(j^*)} } \{ 1, \ldots, 2^{nr_{k,m}} \}    \times \Xc^n &\to \Xc^n \\
 ( i_m )_{m \in \mathcal{I}_{(k)|(j^*)}} ,  \xv^n_{\js} & \mapsto \hat \xv^n_{\k|\js}
\end{align}
\end{subequations} 
\end{itemize}
\end{definition}

\begin{definition}[Achievable storage and transmission rates for a SMRA code] 
\label{def:SMRA_achievablerate}
The \emph{probability of error} for a SMRA code is defined as 
\begin{align}
P_{error}^n& = \P \left( \underset{1 \leq k \leq L}{\cup}  \left(\underset{j^* \in \Jsetk}{\cup} \left(\hat \xv^n_{\k|\js} \neq \xv^n_{\k} \right) \right) \right)
\end{align}

A tuple of rates $((\Sk,( \Rkjs )_{j^* \in \Jsetk })_{1\le k\le L})$ is said to be \emph{achievable for SMRA} if there exists a sequence of $( (2^{n\Sk}, ( 2^{n \Rkjs} )_{j^* \in \Jsetk } )_{1 \le k \le L},n)$ codes such that $  \lim_{n\to +\infty} P_{error}^n =0$.

\end{definition}


The main novelty of the above definition resides in the combination of two encoding mappings: a standard offline encoder that produces the coded source descriptions, and a novel online extractor that can only extract a part of the coded descriptions. 
The online extractor is a very simple operation introduced because re-encoding is not desirable in massive access to data.
The above definition suggests that the encoder and the extractor should be jointly designed in order to minimize both storage $\Sk$ and transmission $\Rkjs$ rates involved in the definition.

The problem studied in this paper can then be restated as follows: what are the minimum storage $\Sk$ and transmission $\Rkjs$ rates that can be achieved by SMRA according to the above definition? 
For sake of clarity, a summary table of the notations introduced in this Section is proposed in Table.~\ref{table:notations}.

\begin{table*}[h]
\begin{center}
  \begin{tabular}{|l|l|}
\hline
$L$ & Number of sources\\ \hline
$l$ & Index of an arbitrary source, with $1\leq l \leq L$\\ \hline
$k$ & Index of the source requested by the client, with $1\leq k \leq L$ \\ \hline
$\Jsetk$ & Set of possible previously requested source indices \\ \hline
$j$ & Index of an arbitrary previously requested source, with $j \in \Jsetk$ \\ \hline
$j^*$ & Index of the source actually requested by the client at previous instant, with $j^* \in \Jsetk$ \\ \hline
$S_{(k)}$ & Storage rate cost for coding the source of index $k$\\ \hline
$R_{(k)|(j*)}$ & Transmission rate cost for coding the source of index $k$ with source $j^*$ in memory\\ \hline
  \end{tabular}
      \caption{Notations used for SMRA problem description.}
  \label{table:notations}
  \end{center}
\end{table*}


%
\section{Revisiting state of the art results in the SMRA framework}\label{sec:soa}


SMRA can be seen as a problem of source coding with side information available at the decoder only~\cite{Slepian73IT,Wyner74IT}.
This is why existing works in the literature may give partial answers regarding the optimal transmission and storage performance of SMRA. 
The objective of this section is to review these works and to restate their results according to the definition of the SMRA coding scheme. 

\subsection{Source coding with multiple side informations}
Sgarro~\cite{sgarro77IT} introduces the problem of source coding with multiple decoders, where each decoder has access to a different side information. 
With our notations (see Table.~\ref{table:notations}), the framework of~\cite{sgarro77IT} considers that the source $\Xk$ has to be transmitted to $|\Jsetk |$ decoders with different side informations $ \Xj \in \Jsetk$.
Nevertheless, Sgarro considers only one encoding mapping $h^{\text{off}}_{\k}$ which means that the same index is transmitted to all decoders. 
Restated in our context, the results of~\cite{sgarro77IT} for i.i.d. sources hence give 
\begin{equation}\label{eq:resultworst}
\Sk = \max_{j \in \Jsetk} H \left(  \Xk | \Xj \right) .
\end{equation}
Further, since no extractor is considered in~\cite{sgarro77IT}, the above result leads to 
\begin{equation}\label{eq:resultworst_rate}
\Rkjs= \displaystyle \max_{j \in \Jsetk} H \left(  \Xk | \Xj \right).
\end{equation}

These rate expressions may also be derived from other source coding problems. 
In particular, Draper and Martinian~\cite{Draper07} already proposed an analysis of FTV from an information-theoretic perspective. 
Nevertheless, they do not distinguish between storage and transmission, and they only introduce a single encoding mapping as in~\cite{sgarro77IT}. 
This is why, in our context, the results of~\cite{Draper07} (derived for i.i.d. sources) can also be restated as~\eqref{eq:resultworst}. 
The results of~\cite{sgarro77IT} and~\cite{Draper07} inspired practical solutions for FTV, see~\cite{Cheung_G_2011_tip_int_ssmvurfs} for instance.

In~\eqref{eq:resultworst} and \eqref{eq:resultworst_rate}, both the storage and the transmission rates derive from the worst possible side information $\Xj$.
These results seem reasonable for the storage rate since the offline encoder does not know which of the sources $\Xj \in \Jsetk$ will have been previously requested by the client.
However, we will show that the knowledge of the current and previous clients requests can be exploited during data transmission in order to send the data at a rate lower than the worst case.

\subsection{Source coding with side information and unknown statistics}
Another problem related to SMRA is source coding with side information at the decoder when the statistics between the source and the side information (\emph{i.e.} their joint probability distribution) are known nor by the encoder nor by the decoder.
In~\cite{Draper04,yang10IT}, in order to deal with the lack of knowledge on the statistics, the encoder successively transmits some pieces of information until the decoder indicates via a feedback link that it was able to successfully decode the message. 
In our context, the framework of~\cite{Draper04,yang10IT} considers that each $\Xj \in \Jsetk$ defines one possible joint probability distribution $P(\Xk,\Xj)$ between the current source $\Xk$ and the previous request. 
However, it is worth noting that our setting does not need a feedback link, since, during the on-line phase, the encoders knows the index of the previously requested source (contrary to the off-line phase).
The results of~\cite{Draper04,yang10IT} indicate that the SMRA coding scheme may be able to achieve
\begin{align}\label{eq:Rfeedback}
 \Rkjs & = H \left(  \Xk | \Xjs \right)
\end{align}
for i.i.d.~\cite{Draper04} and stationary~\cite{yang10IT} sources. 
Compared to~\cite{sgarro77IT,Draper07},~\eqref{eq:Rfeedback} shows that it may be possible to reduce the transmission rate to the conditional entropy of the current request knowing the true previous one. 
Note that the rate expression of~\eqref{eq:Rfeedback} is only a reinterpretation of the results of~\cite{Draper04} in our context. 
In particular,~\cite{Draper04} considers only one encoder, and derives only the transmission rate which is shown to vary with the true source statistics. 

The rate expressions~\eqref{eq:resultworst} and~\eqref{eq:Rfeedback} give insights on the storage and transmission rates that can be achieved in the context of SMRA. 
In the following, we first consider i.i.d. sources for simplicity and we show that the storage rate~\eqref{eq:resultworst} and the transmission rates~\eqref{eq:Rfeedback} can be achieved together and are optimal for the SMRA coding scheme of Definition~\ref{def:SMRAcode}. 
Although we exactly retrieve the expressions~\eqref{eq:resultworst},~\eqref{eq:Rfeedback}, deriving the proofs from the beginning permits to introduce a formalism that will be useful to extend the analysis to other cases of interest in practical situations.
 In the remaining of the paper, we provide extensions to non i.i.d., non stationary sources, to unknown source statistics, and to the case where not only one, but all the previous requests are stored in the client's memory.

\section{Optimal transmission and storage rates for SMRA}
\label{sec:RSbounds:iid}


\subsection{Discrete i.i.d. sources, known statistics}\label{subsec:iid known stats}
We first derive the storage and transmission rates for discrete i.i.d. sources, assuming that all the source statistics are known. The proof of Theorem~\ref{th:iid_stat_known} introduces the concepts that allow the generalization to general sources (Section \ref{sec:RSbounds:gen}), without source statistic knowledge (Sections \ref{sec:iid_stat_unknown} and \ref{sec:RSbounds:gen}), lossy compression (Section \ref{sec:lossy}), and unbounded memory  (Section \ref{sec:RSboundsMRA:iid}).

\begin{theorem} [SMRA with previously requested source stored at the decoder]
Let $\{\Xl\}_{1\le l \le L}$ be a set of $L$ discrete  i.i.d. sources (see Definition~\ref{def:iidSources}). 
The  tuple of rates $((\Sk,( \Rkjs )_{j^* \in \Jsetk })_{1\le k\le L})$ is achievable for SMRA (see Definition~\ref{def:SMRA_achievablerate}) if and only if $\forall k \in \llbracket 1,L \rrbracket, \forall j^{*} \in \Jsetk$
\begin{subequations}\label{eq:RS}
\begin{align}
\Rkjs &\ge H \left(  \Xk | \Xjs \right)  \\
\Sk&\ge \displaystyle \max_{j \in \Jsetk} H \left(  \Xk | \Xj \right)
\end{align}
\end{subequations}
provided  that all statistics of the sources  are known by both the encoder and the decoder, and that $\Mk=|\Jsetk|$ is finite.
\label{th:iid_stat_known}
\end{theorem}

\textit{Remark:}
The  previously requested source $\Xjs$ can be seen as a side information partially available at the encoder and perfectly available at the decoder.
Indeed, during the offline phase, the encoder only knows the set of possible side informations $\Xj$, $j\in\Jsetk$, and their realizations. 
Nevertheless, during the online phase, the extractor has access to the exact index $\js$ of the previously requested source available at the decoder as side information.

\textit{Remark:} When the index $0$ of the root node belongs to $\Jsetk$, \emph{i.e.},  when the source with index $k$ can be directly accessed, the source needs to be stored at its entropy $ H (  \Xk)$ from \eqref{eq:RS}. However, the source can be transmitted at a lower rate, if $j^* \ne 0$.

\begin{proof} \textbf{Achievability}. In this part, we show the existence of a sequence of SMRA codes that achieves the tuple of rates  \eqref{eq:RS}. In what follows, we consider a specific source index $k$. This source can be requested after one other source has been requested and stored at the decoder. The index of the previously requested source belongs to the set $\Jsetk$.

{\em Source ordering.} The first step consists in ordering the set of possible previous requests $\Jsetk$. More precisely, for each source index $k$, let 
\begin{subequations}\label{eq:orderingMap}
\begin{align}
\sigA_{(k)}: \Jc &\to \llbracket 1, \Mk \rrbracket \\
j & \mapsto \sigA_{(k)}(j)
\end{align}
\end{subequations}
be an ordering mapping of the source indexes in $\Jsetk$ such that 
\begin{align}
\!\!\!\!\!  H (  \Xk | X_{(\sigA_{(k)}^{-1}(1))} )  \le ... \le & H (  \Xk | X_{(\sigA_{(k)}^{-1}(m))}) \nonumber \\ 
 & \le ... \le  H (  \Xk | X_{(\sigA_{(k)}^{-1}(\Mk))} ), 
\label{eq:orderSourcePreviousRequest}
\end{align}
where $m$ stands for the order of a particular  source index $j$: $m=\sigA_{(k)}(j)$.

{\em Remark on source ordering.} 
Note that the condition  \eqref{eq:orderSourcePreviousRequest} is weak as it only depends on the conditional entropies and can always be obtained.  In particular, this condition does  not imply that $\Xk\to X_{(\sigA^{-1}(\Mk))} \to  X_{(\sigA^{-1}(m))} \to X_{(\sigA^{-1}(1))}$ is a Markov chain and forms a degraded model. 

{\em Notation simplification.} In what follows, we consider a specific source index $k$ and omit this index to lighten notations.

{\em Coding scheme seen from the offline encoder point of view.} As depicted in Figure~\ref{fig:CodingSchemeForProof}, before sending $\xv^n$, the encoder knows that another sequence (possibly $\xv^n_{\j}$ with $j \in \Jc$) was sent to the decoder and stored in its memory. 
This sequence was chosen by the client from the set of possible sequences $\{\xv^n_{\j}\}_{j\in\Jc}$.
The offline encoder knows the set of possible sequences $\{\xv^n_{\j}\}_{j\in\Jc}$, but does not know which sequence of this set will be available at the decoder, since the encoding is performed offline. 
Due to similarities with the Slepian Wolf problem \cite{Slepian73IT}, we refer to the set of sequences $\{\xv^n_{(j)}\}_{j\in\Jc}$  as 
the  set of possible side informations in the following.
 
 The idea of the proof is to first partition the space $\Xc^n$ into $2^{nr_{1}}$ bins, and then to further partition each bin into $2^{nr_{2}}$ bins. The process is repeated $M$ times and this creates an embedded random binning.

\begin{figure}[h]
\begin{center}
\includegraphics[width=0.7\linewidth]{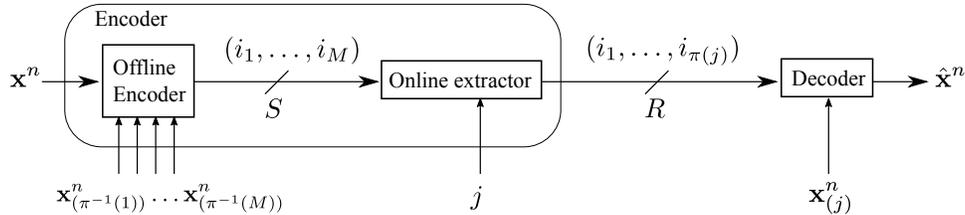}
\end{center}
\caption{Coding scheme seen from the offline encoder point of view. $j\in \Jc$ is a possible previous request available at the decoder. $\sigA(j)$ is the order of $j$ according to \eqref{eq:orderSourcePreviousRequest}.  }
\label{fig:CodingSchemeForProof}
\end{figure}

{\em Random code generation} is performed recursively. First, assign every $\xv^n \in \Xc^n$ to one of $2^{nr_{1}}$ bins independently according to a uniform distribution on $\lb 1,2^{nr_{1}}\rb$. The binning mapping is defined as $f_{1}(\xv^n) = i_{1}$, where $i_{1}$ is the index of the bin assigned to $\xv^n$. Let the bin $\Bc_{i_{1}}=f_{1}^{ -1}(i_{1})$ denote the preimage of $i_{1}$  under the map $f_{1}$.

At the second stage, for each bin index ${i_{1}} \in \lb 1,2^{nr_{1}}\rb$, assign every  $\xv^n \in \Bc_{i_{1}}$  independently according to a uniform distribution on $\lb 1,2^{nr_{2}}\rb$. The binning mapping is defined as $f_{2}(\xv^n) = (i_{1},i_{2})$, where $i_{1}$ and $i_{2}$ are the indexes assigned to $\xv^n$ at the first and second stage respectively. Let the bin $\Bc_{i_{1},i_{2}}=f_{2}^{-1}(i_{1},i_{2})$ denote the preimage of $ (i_{1},i_{2})$  under the map $f_{2}$.

Repeat the sub-partitioning up to stage $M$. At the last stage, the mapping is $f_{M}(\xv^n) = (i_{1},i_{2},...,i_{M})$. Finally, the binning is defined by a set of mappings $\{f_m\}_{1\le m\le M}$, and  revealed to both the encoder and the decoder.

\begin{figure}[h]\begin{center}
\includegraphics[width=0.6\linewidth]{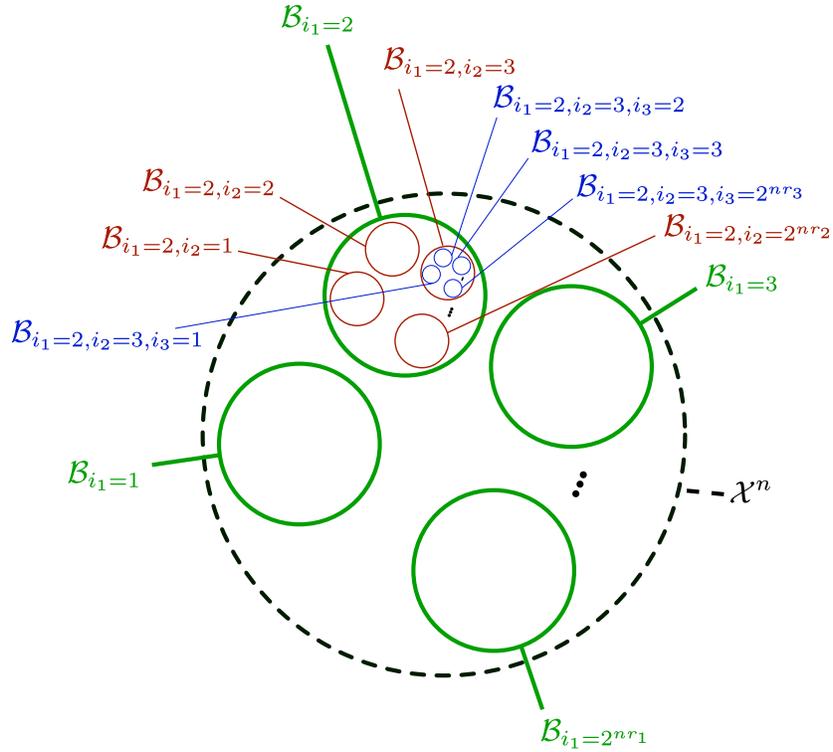}
\end{center}
\caption{Random code generation of one source: embedded random binning}\label{fig:bins}\end{figure}

{\em Encoding} is performed in two parts, see Definition~\ref{def:SMRAcode} and Figure~\ref{fig:CodingSchemeForProof}. The {\em offline encoder} computes a compressed version of the source sequence $\xv^n$ as a sequence of bin indexes:
$$ h^{\text{off}} (\xv^n, ( \xv^n_{\j}  )_{j \in \Jc } ) =  f_{M}(\xv^n) = (i_{1},i_{2},...,i_{M}), $$
where there are as many bin indexes as the number of possible previous requests. $ (i_{1},i_{2},...,i_{M})$ corresponds to the index sequence  of the bin to which $\xv^n$ belongs at the $M^{th}$ and last sub-partitioning. This index sequence is stored in the server.

When the {\em  online extractor} receives the request and, if the index of the source available at the decoder is $j$, it extracts the $\sigA(j)$ first bin indexes from the stored sequence of bin indexes:
$$ h^{\text{on}}_{|(j)} (i_{1},i_{2},...,i_{M}) = (i_{1},i_{2},...,i_{\sigA(j)})$$ 
and sends it to the client. 

{\em Typicality}  The set of jointly typical sequences, denoted $A_{\epsilon}^{(n)} (X, X_{\j})$, is defined as 
\begin{align}
 A_{\epsilon}^{(n)} (X, X_{\j}) &=   \Big\{  \left( \xv^{n}, \xv_{\j}^{n} \right) \in \Xc^{n} \times \Xc^{n}_{\j}: \nonumber\\ 
&  \left| - \frac{1}{n} \log P \left( \xv^{n}| \xv_{\j}^{n} \right) - H(X|X_{\j}) \right| < \epsilon \Big \} \label{eq:jointtyp}
\end{align}
Note that this definition of joint typicality differs from \cite[Section 7.6]{cover2006b}. We further denote $A^n_\epsilon (X |  \xv^n_{\j})$, the set of $\xv^n$ sequences jointly typical with $\xv^n_{\j}$. The cardinality of this set can be upper bounded by
\begin{align}
  |A^n_\epsilon (X |  \xv^n_{\j})| & \le
   2^{n\Large(H(X| X_{\j}) + \epsilon \Large)}
  \label{eq:boundtypical}
\end{align}
 since:
\begin{align}
  1 &\ge 
  \sum_{\mathclap{\xv'^n  \in A^n_\epsilon (X |  \xv^n_{\j})}}  P(\mathbf{x'}^{n}|\mathbf{x}_{(j)}^{n}) \ge |A^n_\epsilon (X |  \xv^n_{\j})| 
  . 2^{-n\Large(H(X| X_{\j}) + \epsilon \Large)}
\end{align}
where the last equality follows from \eqref{eq:jointtyp}. 

{\em Decoding} Given the received index sequence $(i_{1},...,i_{\sigA(j)})$ and the side information $\xv_{\j}$, declare $g((i_{1},...,i_{\sigA(j)}),\xv_{\j}) = \hat \xv^n = \xv^n$ if there is a unique pair of sequences $(\xv^n, \xv^n_{\j})$ such that $f_{\sigA(j)}(\xv^n) =(i_{1},..., i_{\sigA(j)})$ and $(\xv^n, \xv^n_{\j})$ is jointly typical, denoted $(\xv^n, \xv^n_{\j}) \in A^n_\epsilon(X, X_{\j})$.
Otherwise, declare an error. 

{\em Probability of error}.  We now compute the error probability for each possible previously requested source $\Xj$, where $ j \in \Jc$. 
Let $(\Xv^n,\Xv^n_{\j})\sim P( \xv^n,\xv^n_{\j} )$. We define the events
\begin{align}
E_0&= \{ (\Xv^n,\Xv^n_{\j}) \notin A^n_\epsilon(X, X_{\j})\} \\
E_j&= \{ \exists \xv'^n \neq \Xv^n: f_{\sigA(j)} (\xv'^n) =  f_{\sigA(j)} (\Xv^n)   \nonumber \\
& \ \ \ \ \ \ \ \ \ \ \ \  \mbox{ and } (\xv'^n,\Xv^n_{\j}) \in A^n_\epsilon(X, X_{\j})\}, \ \ \ \forall j \in \Jc
\end{align}
Indeed, there is an error if $(\Xv^n,\Xv^n_{\j})$ is not typical or if there is another typical sequence pair in the same bin. Then, by the union bound, the error probability is
\begin{align}
P_{error}^n &= \P ( E_0 \bigcup_{j\in \Jc}  E_j) \le \P ( E_0 ) + \sum_{j\in \Jc} \P ( E_j ) 
\end{align}

First, consider $E_0$. By the Law of Large Number applied to the i.i.d. random source $-\log P(X|X_{\j})$, we have that 
\begin{align}
-\frac{1}{n}\sum_{i=1}^n \log P(X_i|X_{\j,i}) &\to  H(X|X_{\j}) \mbox{ in probability}.
\end{align}
Therefore, using similar arguments as for the Asymptotic equipartition property (AEP) \cite[Th. 3.1.2]{cover2006b}, we have $\P ( E_0 ) \to 0$ as $n\to\infty$ and hence, $\forall \epsilon>0$, for n sufficiently
large, $\P(E_0) < \epsilon$.

We now compute $E_j$, $\forall j$
\begin{align}
\P ( E_j )&  = \sum_{{(\xv^n,\xv^n_{\j})  \in  \Xc^n\times \Xc^n_{\j} } } P ( \xv^n,\xv^n_{\j} ) ...&\\
& \P 
\Big( \exists \xv'^n \neq \xv^n: f_{\sigA(j)} (\xv'^n) =  f_{\sigA(j)} (\xv^n)  \mbox{ and } (\xv'^n,\xv^n_{\j}) \in A^n_\epsilon
\Big) \nonumber \\
&  \le \sum_{\mathclap{(\xv^n,\xv^n_{\j})} } P ( \xv^n,\xv^n_{\j} ) \sum_{\mathclap{\substack{\xv'^n \neq \xv^n \\ (\xv'^n,\xv^n_{\j}) \in A^n_\epsilon}}}
\P \Big( f_{\sigA(j)} (\xv'^n) =  f_{\sigA(j)} (\xv^n)   \Big) \label{eq:Ej_1}\\
&\le \sum_{\mathclap{(\xv^n,\xv^n_{\j})} } P ( \xv^n,\xv^n_{\j} )  . \big| A^n_\epsilon (X |  \xv^n_{\j} )\big| . 2^{ -n \big( r_{1}+  ... +r_{\sigA(j)}\big)}  \label{eq:Ej_2}\\
&\le 2^{ -n \big( r_{1}+  ... +r_{\sigA(j)}\big)}     .2^{n\Large(H(X| X_{\j}) + \epsilon \Large) }\label{eq:Ej_3}
\end{align}
The probability in \eqref{eq:Ej_1} is computed over all realizations of the random mapping $f_{\sigA(j)}$. The inequality in \eqref{eq:Ej_1} follows from the fact that there might be many $\xv'^n$ satisfying the condition (of the second sum) and from the fact that these events are not disjoint (union bound). For~\eqref{eq:Ej_2}, the indexes are chosen according to a uniform distribution, therefore the probability equals $ 2^{ -n \big( r_{1}+  ... +r_{\sigA(j)}\big)}$ and $A^n_\epsilon (X |  \xv^n_{\j})$ is the set of $\xv^n$ sequences jointly typical with $\xv^n_{\j}$. Finally, since the upper bound of $ |A^n_\epsilon (X |  \xv^n_{\j})| $ \eqref{eq:boundtypical} and the probability $ 2^{ -n ( r_{1}+  ... +r_{\sigA(j)})}$ are independent of $\xv^n,\xv^n_{\j}$ and $\xv'^n$, we get  \eqref{eq:Ej_3}.

Then, \eqref{eq:Ej_3} goes to 0 if $r_{1}+  ... +r_{\sigA(j)} > H(X| X_{\j}) + \epsilon$. Hence, $\forall j \in \Jc$ and for sufficiently large $n$, $\P(E_{j}) < \epsilon$.
Thus the average probability of error $P_{error}^n$
is less than $ (M+1) \epsilon$. So, there exists at least one code $\{f_{\sigA(j)}\}_{j}$
with probability of error less than
 $ (M+1) \epsilon$. Thus,  if $M$ is finite, we can construct a sequence of codes (indexed by $n$) with vanishing $P_{error}^n $.

In other words, for any previously requested source $j \in \Jc$, there exists a code with vanishing error rate and compression rate $r_{1}+  ... +r_{\sigA\j}$, provided that $r_{1}+  ... +r_{\sigA\j} \ge H(X| X_{\j}) $ and that $M$ is finite. 
Finally, to be able to serve any request, the compression scheme needs to prepare a code for any $j \in \Jc$.
  Therefore, the storage requires  $S \ge \max_{j \in \Jc} H \left(  X | \Xj \right)$.
  Moreover, if the actual previous request is $j^*$, the transmission rate $R$ needs to satisfy $R =r_{1}+  ... +r_{\sigA\js} \ge H(X| X_{\js})$,  which completes the achievability proof.


  %
\textbf{\textit{Converse: outer bound.}}\\ 
$\bullet \  \Rkjs \ge H \left(  \Xk | \Xjs \right)$. Proof by contradiction. If   the encoder knows perfectly the realization of the side information $\xv^n_{\js}$, and if $\Rkjs < H \left(  \Xk | \Xjs \right)$, then, by a conditional version of the lossless (single) source coding theorem \cite[Sec. 5.2]{yeung2010}, the decoder fails to recover the source with probability $1$ as $n\to \infty$. Therefore, if the encoder (compression part) does not know $\xv^n_{\js}$, the decoder will also fail.\\
$\bullet \ \Sk \ge \displaystyle \max_{j \in \Jc} H \left(  \Xk | \Xj \right)$. Proof by contradiction.\\ 
If $\Sk< \max_{j \in \Jc} H \left(  \Xk | \Xj \right)$ , then with non zero probability, the side information might be $X_{( \bar\jmath)}$, where $ \bar\jmath = \arg \max_{j \in \Jc} H \left(  \Xk | \Xj \right)$.  From the converse of the conditional version of the lossless source coding theorem \cite[Sec. 5.2]{yeung2010}, this source can not be recovered at the decoder.  
\end{proof}

\subsection{Non \emph{i.i.d.} sources, known statistics}
In practical situations, sources such as the frames in FTV, or measurements collected by the nodes of a sensor network, present internal non-stationary correlation that needs to be exploited.
Such sources correspond to non i.i.d. sources, see Definition~\ref{def:generalSources}.
As for the i.i.d. setup, a user wants to access a source $\Xk$ among a set of $L$ sources $\{\Xl\}_{1\le l \leq L}$.
However, in this case, since the sources are not \emph{i.i.d.}, the joint probability distribution $P(\xv_{(1)}^n, \cdots, \xv_{(L)}^n) $ cannot be factorized as in~\eqref{eq:factoriid}.

For two non i.i.d. sources $X_{(l)}$ and $X_{(j)}$, the \emph{spectral conditional entropy} $ \Hs \left(  \Xl | \Xj \right)$ is defined from a  $\lim\sup$ in probability.
The $\lim\sup$ in probability of a sequence $\{A_n\}_{n=1}^{+\infty}$ of random variables is denoted as $ \plim A_n$ and is given by
\begin{equation}
 \plim A_n = \inf\left\{ \alpha | \lim_{n \rightarrow +\infty} \P(A_n > \alpha) = 0 \right\} .
\end{equation}
The spectral conditional entropy $ \Hs \left(  \Xl | \Xj \right)$ is then defined as~\cite[Chapter 7]{han2003b}
\begin{equation}\label{eq:defHs}
  \Hs\left(  \Xl | \Xj \right) = \plim \frac{1}{n} \log \frac{1}{P(\Xv_{(l)}^{n}|\Xv_{(j)}^{n})}
\end{equation}
In~\eqref{eq:defHs}, the vectors $\Xv_{(l)}^{n},\Xv_{(j)}^{n}$ are random vectors distributed according to the joint probability distribution $P(\xv_{(l)}^{n},\xv_{(j)}^{n}) $ that can be obtained from  $P(\xv_{(1)}^n, \cdots, \xv_{(L)}^n)$.
In the above definition, the $\lim\sup$ in probability is required because general sources are not necessarily ergodic, and hence the random variable $\frac{1}{n} \log \frac{1}{P(\Xv_{(l)}^{n}|\Xv_{(j)}^{n})}$ does not necessarily converge to a unique value.
It can be shown that if the sources $\Xl $ and $\Xj$ are i.i.d., then the spectral conditional entropy reduces to the standard definition of entropy, that is $\Hs \left(  \Xl | \Xj \right) =  H\left(  \Xl | \Xj \right)$.

From the definition of the spectral conditional entropy in~\eqref{eq:defHs}, the rate-storage region can be obtained for non i.i.d. sources by restating Theorem~\ref{th:iid_stat_known} as follows.

\begin{theorem} 
Let $\{\Xl\}_{1\le l \le L}$ be a set of $L$ general sources (see Definition \ref{def:generalSources}). 
The  tuple of rates $((\Sk,( \Rkjs )_{j^* \in \Jsetk })_{1\le k\le L})$ is achievable for SMRA (see Definition~\ref{def:SMRA_achievablerate})  if and only if $\forall k \in \llbracket 1,L \rrbracket, \forall j^{*} \in \Jsetk$
\begin{subequations}\label{eq:RS_noniid}
\begin{align}
\Rkjs &\ge \Hs \left(  \Xk | \Xjs \right)\\[.1cm]
\Sk&\ge \displaystyle \max_{j \in \Jsetk} \Hs \left(  \Xk | \Xj \right),
\end{align}
\end{subequations}
provided  that all statistics of the sources  are known by both the encoder and the decoder, and that $\Mk$, the  size of the set of possible  previous requests $\Jsetk$ is finite.
\label{th:noniid_stat_known}
\end{theorem}

\begin{proof} \textbf{Achievability.} In this part, we show the existence of a sequence of codes that achieves the set of rates \eqref{eq:RS_noniid}. 
As for the i.i.d. case, the first step consists in ordering the set of possible previous requests. For each source index $k$, let 
\begin{subequations}\label{eq:orderingMapni}
\begin{align}
\sigA: \Jsetk &\to \llbracket 1, \Mk \rrbracket \\
j & \mapsto \sigA(j)
\end{align}
\end{subequations}
be an ordering mapping of the source indexes in $\Jsetk$ such that
\begin{align}\notag
\!\!\!\!\!  \Hs (  \Xk | X_{(\sigA^{-1}(1))} )  \le ... \le & \Hs (  \Xk | X_{(\sigA^{-1}(m))}) \\ \notag & \le ... \le  \Hs (  \Xk | X_{(\sigA^{-1}(\Mk))} ).
\label{eq:orderSourcePreviousRequestnoniid}
\end{align}

 We keep the same notations and conventions as for the proof of Theorem~\ref{th:iid_stat_known}.
 In particular, we omit the index $(k)$ of the source to be encoded.
 We also rely on the same \textit{random code generation} and \textit{encoding} as for the proof in the \emph{i.i.d.} case.
 Indeed, these two steps do not depend on the \emph{i.i.d.} or non \emph{i.i.d.} nature of the sources, except for the choice of the rates $r_1, \cdots, r_M$, which will be addressed later in the proof.
 However, the \emph{decoding} as well as the \emph{error probability analysis} differ from the \emph{i.i.d.} case, as we now describe.
 
{\em Decoding.} For general sources, define the set $T^n_\epsilon(X, X_{\j})$ as
 \begin{equation}\label{eq:Adef}
  T_{\varepsilon}^{(n)}(X,X_{\j}) = \left\{ (\xv^{n},\xv_{\j}^{n}) \left| \frac{1}{n} \log \frac{1}{P(\mathbf{x}^{n}|\xv_{\j}^{n})} < \sum_{i=1}^{\sigA(j)} r_i - \varepsilon \right. \right\} .
 \end{equation}
Given the received index sequence $(i_{1},...,i_{\sigA(j)})$ and the side information $\xv_{\j}^n$, declare $\hat{\xv}^n = \xv^n$ if there is a unique pair of sequences $(\xv^n, \xv_{\j}^n)$ such that $f_{{\sigA(j)}}(\xv^n) =(i_{1},...,i_{\sigA(j)})$ and $(\xv^n, \xv_{\j}^n) \in T^n_\epsilon(X,X_{\j})$. Otherwise, declare an error.

 {\em Probability of error.}
  We define the events
\begin{align}
E_0&= \{ (\Xv^n,\Xv_{\j}^n) \notin T^n_\epsilon(X, X_{\j})\} \\
E_j&= \{ \exists \xv'^{n} \neq \Xv^n: f_{\sigA(j)} (\xv'^{n}) =  f_{\sigA(j)} (\Xv^n)   \nonumber \\
& \ \ \ \ \ \ \ \ \ \ \ \  \mbox{ and } (\xv'^{n},\Xv_{\j}^{n}) \in T^n_\epsilon(X, X_{\j})\}, \ \ \ \forall j \in \Jc
\end{align}
From the union bound, the error probability is given by
\begin{align}\label{eq:pen_noniid}
 P_{\text{error}}^n &= \P ( E_0 \bigcup_{j\in \Jc}  E_j) \le \P ( E_0 ) + \sum_{j\in \Jc} \P ( E_j ) 
\end{align}

We first consider the error events $E_j$, $\forall j \ge 1$. The error probability $\P(E_j)$ can be derived as for the \emph{i.i.d.} case as 
\begin{align}
\P ( E_j )&  = \sum_{{(\xv^n,\xv_{\j}^n)  \in  \Xc^n\times \Xc^n_{\j} } } P ( \xv^n,\xv_{\j}^n ) ...&\\
& \P 
\Big( \exists \xv'^n \neq \xv^n: f_{\sigA(j)} (\xv'^n) =  f_{\sigA(j)} (\xv^n)  \mbox{ and } (\xv'^n,\xv_{\j}^n) \in T^n_\epsilon
\Big) \nonumber \\
&  \le \sum_{\mathclap{(\xv^n,\xv_{\j}^n)} } P ( \xv^n,\xv_{\j}^n ) \sum_{\mathclap{\substack{\xv'^n \neq \xv^n \\ (\xv'^n,\xv_{\j}^n) \in T^n_\epsilon}}}
\P \Big( f_{\sigA(j)} (\xv'^n) =  f_{\sigA(j)} (\xv^n)   \Big) \label{eq:Ej_1niid}\\
&\le \sum_{\mathclap{(\xv^n,\xv_{\j}^n)} } P ( \xv^n,\xv_{\j}^n)  . \big| T^n_\epsilon (X |  \xv_{\j} )\big| . 2^{ -n \left( \underset{i=1}{\overset{\sigA(j)}{\sum}} r_i \right)}  \label{eq:Ej_2_niid}
\end{align}
 where the set  $T_{\varepsilon}^{(n)}(X|\xv_{\j}^n)$ is defined as
  \begin{equation}\label{eq:Adef_niid}
  T_{\varepsilon}^{(n)}(X|\xv_{\j}^n) = \left\{ \xv^{n} \left| \frac{1}{n} \log \frac{1}{P(\mathbf{x}^{n}|\xv_{\j}^{n})} < \sum_{i=1}^{\sigA(j)} r_i - \varepsilon \right. \right\} .
 \end{equation}\color{black}
 Here, $| T^n_\epsilon (X |  \xv_{\j} )\big|$ can be bounded using the fact that
  \begin{equation}\label{eq:boundA}
  1 \geq \hspace{-0.3cm} \sum_{\mathbf{x'}^{n} \in T_{\varepsilon}^{(n)}(X|\mathbf{x}^n_{(j)})} \hspace{-0.6cm} P(\mathbf{x'}^{n}|\mathbf{x}_{(j)}^{n}) \geq |T_{\varepsilon}^{(n)}(X|\mathbf{x}^n_{(j)})| \exp\left(-n \sum_{i=1}^{\sigA(j)} r_i +n\varepsilon\right)
 \end{equation}
where $P(\mathbf{x'}^{n}|\mathbf{x}_{(j)}^{n}) \geq \exp\left(-n\sum r_i +n\varepsilon \right) $ comes from the definition of $T_{\varepsilon}^{(n)}(X|\xv_{\j}^n)$ in~(\ref{eq:Adef_niid}).
Equation~\eqref{eq:boundA} gives 
\begin{equation}
 |T_{\varepsilon}^{(n)}(X|\mathbf{x}^n_{(j)})| \leq \exp\left(n\sum_{i=1}^{\sigA(j)} r_i -n\varepsilon\right)
\end{equation}
and as a consequence the error probability $\P(E_j)$ becomes
\begin{equation}
 \P ( E_j ) \leq \exp(-n\epsilon) .
\end{equation}
This gives $\lim_{n \rightarrow \infty} \P(E_j) = 0$.

Now, according to the definition of $T_{\varepsilon}^{(n)}(X,X_{\j})$ in~\eqref{eq:Adef_niid}, the error probability $\P(E_0)$ can be expressed as
\begin{equation}
 \P(E_0) = \P\left(\frac{1}{n} \log \frac{1}{P(\Xv^{n}|\Xv_{(j)}^{n})} \geq \sum_{i=1}^{\sigA(j)} r_i+ \varepsilon \right)
\end{equation}
 If $\underset{i=1}{\overset{\sigA(j)}{\sum}} r_i \geq \Hs \left(  X | \Xj \right)$, then from the definition of $ \Hs \left(  X | \Xj \right)$ in~(\ref{eq:defHs}), $\P(E_0)$ goes to $0$ when $n$ goes to infinity.

From the analysis of $\P(E_0)$ and $\P(E_j)$, and from the expression of $P_{\text{error}}^n$ in~\eqref{eq:pen_noniid}, there exists at least one code $\{f_{\sigA(j)}\}_j$
with probability of error  less than $ (M+1) \epsilon$. Thus, if $M$ is finite, we can construct a sequence (indexed by $n$) of codes with $P_{\text{error}}^n \to 0$.
 
In other words, for any previously requested source $j \in \Jc$, there exists a code with vanishing error rate and compression rate $r_{1}+  ... +r_{\sigA\j}$, provided that $r_{1}+  ... +r_{\sigA\j} \ge \Hs(X| X_{\j}) $ and that $M$ is finite. 
Finally, to be able to serve any request, the compression scheme needs to prepare a code for any $j \in \Jc$.
  Therefore, the storage requires  $S \ge \max_{j \in \Jc} \Hs \left(  X | \Xj \right)$.
  Moreover, if the actual previous request is $j^*$, the transmission rate $R$ needs to satisfy $R =r_{1}+  ... +r_{\sigA\js} \ge \Hs(X| X_{\js})$,  which completes the achievability proof. 

 \textbf{Converse.} The converse is done by contradiction as for the \emph{i.i.d.} case.
 In particular, the fact that if $\Rkjs < \Hs \left(  \Xk | \Xjs \right)$, the decoder fails to recover the source comes from~\cite[Theorem 7.2.1]{han2003b}.
 
 The same result~\cite[Theorem 7.2.1]{han2003b} shows that if $\Sk< \max_{j \in \Jc} \Hs \left(  \Xk | \Xj \right)$, then with non zero probability, the source indexed by $ \bar \jmath  = \arg \max_{j \in \Jc} \Hs \left(  \Xk | \Xj \right)$ can not be recovered at the decoder.  
\end{proof}

\textit{Remark:} In the proof of Theorem~\ref{th:iid_stat_known} for \emph{i.i.d.} sources, the rate condition $R \ge H(X| X_{\js}) $ comes from the error events $E_j$.
On the opposite, in the proof of  Theorem~\ref{th:noniid_stat_known} for general sources, the rate condition $R \ge \Hs(X| X_{\js}) $ comes from $E_0$.
This is due to the fact that the decoding is different from one case to another.
In the i.i.d. case, the decoding relies on the condition that $(\xv^n, \xv_{\js}^n) \in A^n_\epsilon(X, X_{\js})$, the set of jointly typical sequences. 
But the notion of joint typicality cannot be defined for general sources, and hence in the non i.i.d. case, the decoded sequence must satisfy $(\xv^n, \xv_{\js}^n) \in T^n_\epsilon(X, X_{\js})$.

%

\section{Optimal transmission and storage rates for {universal} SMRA}
\label{sec:RSbounds:gen}

When considering SMRA with real data such as pictures or videos, it is usually difficult to have access to reliable source statistics prior to encoding, since these statistics heavily depend on the considered data. 
This is why this section assumes that the source statistics are unknown. Generally speaking, this means that the coding scheme does not know anything about the joint probability distribution of the sources, nor about their marginals or any of the conditional probability distributions. 
Since the general case may be difficult to characterize, we restrict our attention to two particular cases. Section~\ref{sec:iid_stat_unknown} considers  i.i.d. sources, while Section~\ref{sec:noniid_stat_unknown} considers non i.i.d. sources with joint distribution defined by an unknown parameter vector.   

In both cases, we consider that the source statistics can be inferred from the source realizations. The coding scheme has the same structure as when the source statistics are known, with an offline encoder, an online extractor, and a decoder. The offline encoder now estimates the joint statistics of any possible source pair $(X_{\k},  X_{\j}  )_{j \in \Jsetk } $ in the form of a \emph{learning sequence}. The offline encoder uses this learning sequence together with the source sequences to compress the source $X_{\k}$ into a \emph{ data sequence}. Upon the client's request, the online extractor sends a subsequence of both the learning sequence and the data sequence to the decoder that estimates the requested source. 

The following definition of universal SMRA codes considers variable length coding because the coding mappings and the rates now depend on the realizations of the source sequences $(\xv^n_{\k}, ( \xv^n_{\j}  )_{j \in \Jsetk})$ . In this Definition, the set $\Bc$ is a finite source alphabet with cardinality greater or equal to $2$, and the set $\Bc^+$ is the set of all finite strings from $\Bc$.  In addition, $\forall b \in \Bc^+$, $|b|$ denotes the length of $b$.

\begin{definition}[Universal SMRA   code, storage and transmission rates] 
\label{def:SMRA_universalcode} 
Let $\{ \Xk \}_{1\leq k \leq L}$ be a set of discrete general sources. 
An $n$-length universal SMRA  code consists, for each source $\Xk$, $1\leq k\leq L$, of a set of mappings $\left( h^{\text{off}}_{\k} \right.$,$\left.\left(h^{\text{on}}_{{(k)|(j^*)}},  g_{{(k)|(j^*)}}\right)_{j^* \in \Jsetk}\right)$, where

\begin{itemize}
 \item  $h^{\text{off}}_{\k}$ is the \emph{offline encoder}  that assigns two {sequences of $\Mk=|\Jsetk|$  strings} each (the first one for learning and the second for the data) 
for each possible set of vectors $\left(\xv^n_{\k}, ( \xv^n_{\j}  )_{j \in \Jsetk }\right) \in \Xc^n \times \Xc^{n\times\Mk}$ 
 \begin{align}\notag 
h^{\text{off}}_{\k}: \Xc^n \times \Xc^{n \times \Mk} &\to  (\Bc^+)^{\Mk} \times (\Bc^+)^{\Mk} \\ \notag
\xv^n_{\k}, ( \xv^n_{\j}  )_{j \in \Jsetk } & \mapsto ( \mathbf{i}'_{(j)})_{j \in \Jsetk }, ( \mathbf{i}_{m})_{m \in \Jsetk }
\end{align}


\item  the $h^{\text{on}}_{{(k)|(j^*)}}$ , $j^* \in \Jsetk$, are the $\Mk$  \emph{online sequence extractors}, that extract a string from the sequence of learning strings $ ( \mathbf{i}'_{(j)})_{j \in \Jsetk }$, and a subsequence of strings from the sequence of data strings $ ( \mathbf{i}_{m})_{m \in \Jsetk } $
\begin{align} \notag 
h^{\text{on}}_{{(k)|(j^*)}}: (\Bc^+)^{\Mk} \times (\Bc^+)^{\Mk} &\to  \Bc^+ \\ \notag
 ( \mathbf{i}'_{(j)})_{j \in \Jsetk }, ( \mathbf{i}_{m})_{m \in \Jsetk } & \mapsto  \mathbf{i'}_{\js}, (\mathbf{i}_m )_{m \in \mathcal{I}_{(k)|(j^*)}},  
\end{align}
where $\mathcal{I}_{(k)|(j^*)} \subseteq \{1, \cdots, \Mk\}$.


\item the $g_{{(k)|(j^*)}}$, $j^* \in \Jsetk$ are the $\Mk$ \emph{decoders}  $g_{{(k)|(j^*)}}$, $j^* \in \Jsetk$, that, given the source realization $\xv^n_{\js} $, assign an estimate $\hat \xv^n_{\k|\js}$ to each received learning string and to each subsequence of data strings
\begin{align} \notag 
g_{{(k)|(j^*)}}:    \Bc^+ \times (\Bc^+)^{|\mathcal{I}_{(k)|(j^*)}|}   \times \Xc^n &\to \Xc^n \\ \notag
   \mathbf{i'}_{\js}  ,  ( \mathbf{i}_m )_{m \in \mathcal{I}_{(k)|(j^*)}} ,  \xv^n_{\js} & \mapsto \hat \xv^n_{\k|\js}
\end{align}

\end{itemize}

%
%

The \emph{storage rate} of a universal SMRA code is defined as the number of stored bits per source sample, when encoding a source of finite length $n$. It is the sum of two rates: 
\begin{align}
s_{\k}^{n}(\xv^n_{\k}, ( \xv^n_{\j}  )_{j }) 
& = \frac{1}{n} \left(\sum_{j \in \Jsetk} |\mathbf{i'}_{\j}| + \sum_{m \in \Jsetk} |\mathbf{i}_{m}|\right) \log_2 |\Bc|.
\end{align}
 
The \emph{ transmission rate}  $r_{\k|\js}^{(n)}$ of a universal SMRA code is defined as the number of bits  per source sample sent, when encoding a source of finite length $n$.
\begin{align}
r_{\k|\js}^{n}(\xv^n_{\k}, \xv^n_{\js}  ) 
&= \frac{ 1}{n} \left( |\iv'_{\js}| + \sum_{m \in \mathcal{I}_{(k)|(j^*)}}|\iv_m|\right) \log_2 |\Bc|
\end{align}

\end{definition}

\subsection{Discrete i.i.d. sources, unknown statistics}\label{sec:iid_stat_unknown}

The next Theorem derives the storage and transmission rates for discrete i.i.d. sources when the joint source distribution is neither known at the encoder nor at the decoder. In particular, it shows that this lack of knowledge does not increase the transmission and storage rates, provided a universal code is used. Before deriving this result, a set of i.i.d. sources with unknown statistics is formally defined. This allows to define the notion of achievable rate for universal SMRA code for i.i.d. sources.

\begin{definition}[Discrete i.i.d. source with unknown statistics]
Let $\Xc$ be a discrete alphabet. 
 Let $\Pset$ be a set of joint probability mass functions of dimension $L$ defined over $\mathcal{X}^{L }$.
The set $\{\Xl\}_{1\le l \le L}$ of $L$ discrete i.i.d. sources  is said to be {\em with unknown statistics}  if the joint probability distribution $P$ in (\ref{eq:factoriid}) belongs to the set $\Pset$, and if the set $\Pset$ is known by the encoder and the decoder, but the probability mass function $P$ is unknown.
\label{def:iidSources_stat_unknown}
\end{definition}

\begin{definition}[Achievable storage and transmission rates for i.i.d. sources] 
\label{def:SMRA_universalcode_achievable} 
Let  $\{\Xl\}_{1\le l \le L}$ be a set of $L$ discrete i.i.d. sources with unknown statistics. Given $P$ the unknown probability mass function,
the \emph{probability of error} for a universal SMRA code is defined as 
\begin{align}
P_{error}^n(P)& = \P \left( \underset{1 \leq k \leq L}{\cup}  \left(\underset{j^* \in \Jsetk}{\cup} \left(\hat \Xv^n_{\k|\js} \neq \Xv^n_{\k} \right) \right) \right)
\end{align}
where the event probability is defined with respect to $P$.

A tuple of rates $((\Sk(P),( \Rkjs(P) )_{j^* \in \Jsetk })_{1\le k\le L,P\in \Pset})$ is said to be \emph{universally achievable for SMRA} if there exists 
a  \emph{universal SMRA code} such that $\forall k, j, j^*, P,$
\begin{subequations}\label{eq:cond_iid_stat_unknown}
\begin{align}
&\prlim s_{\k}^{n}(\Xv^n_{\k}, ( \Xv^n_{\j}  )_{j })= \Sk(P)\\
& \prlim  r_{\k|\js}^{n}(\Xv^n_{\k}, ( \Xv^n_{\j}  )_{j })= \Rkjs(P)\\
 &  \lim_{n\to +\infty} P_{error}^n (P)=0
\end{align}
\end{subequations}
where $\prlim$ stands for the limit in probability, and where the event probabilities in the $\prlim$ are defined with respect to $P$.
\end{definition}

\begin{theorem}
Let $\{\Xl\}_{1\le l \le L}$ be a set of $L$ discrete i.i.d. sources with unknown statistics (see Definition~\ref{def:iidSources_stat_unknown}).
The tuple of rates $((\Sk(P),( \Rkjs(P) )_{j^* \in \Jsetk })_{1\le k\le L,P\in \Pset})$ is universally achievable for SMRA (see Definition~\ref{def:SMRA_universalcode_achievable}) if and only if $\forall k \in \llbracket 1,L \rrbracket, \forall j \in \Jsetk, \forall P \in \Pset$,  
\begin{subequations}\label{eq:RS_iid_stat_unknown}
\begin{align}
\Rkjs(P) &\ge H \left(  \Xk | \Xjs \right)  \\
\Sk(P)&\ge \displaystyle \max_{j \in \Jsetk} H \left(  \Xk | \Xj \right)
\end{align}
\end{subequations}
where the entropies are computed with respect to $P$ and provided that $\Mk$, the  size of the set of possible  previous requests $\Jsetk$ is finite.
\label{th:iid_stat_unknown}
\end{theorem}

\begin{proof}
First, compute the joint $p\xkJxj$ and conditional $p\xkxj$ type \cite{MoulinIT07} of each sequence pair $(\xv_{\k}^n,\xv_{\j}^n), \forall j \in \Jsetk$. 
The number of conditional types is upper bounded by $(n+1)^{|\Xc|^2}$  \cite[Th. 11.1.1]{cover2006b}, \cite{MoulinIT07} and therefore requires at most $|\Xc|^2  \log_2 ( n+1) $ bits to be completely determined.
 The conditional type class $T\xkxj$ is the set of all $\xv_{\k}^n$  sequences such that $(\xv_{\k}^n,   \xv_{\j}^n) $ belongs to the joint type class $T\xkJxj$.  From \cite{MoulinIT07}, the cardinality of this set is upper bounded by
\begin{align}
|T\xkxj| & \le 2^{n H(p\xkxj)}
\end{align}
where $H(p\xkxj)$ is the empirical conditional entropy:
\begin{align}
H(p\xkxj) &=  \frac{1}{n} \sum_{i=1}^n \log p\xkxj (x_{\k,i},x_{\j,i})
\end{align}

Second, construct the same code as in Theorem~\ref{th:iid_stat_known} where the true distribution is replaced by the type. 
The learning sequence of the offline encoder $h^{\text{off}}_{\k}$ consists of all conditional types $\{p\xkxj\}_{j \in  \Jsetk}$, which requires $M |\Xc|^2  \log_2 ( n+1) $ bits.
The data sequence is the index sequence computed by the code, which requires $\max_j nH(p\xkxj)$ bits.

When the online extractor receives the request, it sends to the decoder the conditional type $p\xkxjs$ and the subsequence of length $nH(p\xkxjs)$.

The decoder reconstructs the code from the conditional type which leads a vanishing error probability as in Theorem~\ref{th:iid_stat_known}.

 Therefore, the storage and transmission of an $n$-length realization of the source $\Xk$ requires:
\begin{subequations}
\begin{align}
n s_{\k}^{n}(\Xv^n_{\k}, ( \Xv^n_{\j}  )_{j }) &= 
\max_{j \in \Jsetk} nH(p\xkxj) + M |\Xc|^2  \log_2 ( n+1) \\
n r_{\k|\js}^{n}(\Xv^n_{\k}, ( \Xv^n_{\j}  )_{j })& = 
 n H (p\xkxjs)   + |\Xc|^2  \log_2 ( n+1)  
\end{align}
\end{subequations}
By the weak law of large numbers \cite[Th. 11.2.1]{cover2006b} and by continuity of the $\log$ function, the empirical entropies converge to the entropies.
Moreover, if $\Mk$ is finite, then the contribution of the number of types is negligible compared to the contribution of the index as $n \to \infty$. Therefore, the set of achievable rates remains unchanged \eqref{eq:RS}. As for the converse, we use the same converse as in the proof of Theorem \ref{th:iid_stat_known}. Indeed, the encoder in Theorem~\ref{th:iid_stat_unknown} has even less information than in Theorem \ref{th:iid_stat_known} (statistics are unknown).
\end{proof}



%

\subsection{Non i.i.d. parametric sources, unknown statistics}\label{sec:noniid_stat_unknown}
Unfortunately, when the statistics of a non i.i.d. source are unknown, we cannot follow the same reasoning as for i.i.d. sources.
Indeed, the proof of Theorem~\ref{th:iid_stat_unknown} relies on the fact that the encoder can determine the joint type of the sequence pair $(\xv,\xv_{\j}^n)$.
The notion of type is not always defined for non i.i.d. sources since all the symbols of the sequence may statistically depend from each other.
Hence, here, we consider additional assumptions for the non i.i.d. sources, in order to be able to give expressions of the transmission rates and of the storage rate when the sources statistics are unknown.
The assumptions we add are general enough to represent a large class of sources.
The model we consider corresponds to the Sources without Prior (WP-sources) defined in~\cite{dupraz:Tcom14}.

\begin{definition}[WP-sources]\label{def:SPSourcesMRA}
Let $\boldsymbol{\varTheta}$ be a set of vectors of real-valued vectors of length $Q$. 
For all $\boldsymbol{\theta} \in \boldsymbol{\varTheta}$,  denote by $P_{\boldsymbol{\theta}}(\xv_{(1)}^n, \cdots, \xv_{(L)}^n)$ a joint probability mass function of $L$ sequences of length $n$, $(\xv_{(1)}^n, \cdots, \xv_{(L)}^n) \in \mathcal{X}^{L \times n}$.
Assume that the function $\mathbf{t} \mapsto P_{\mathbf{t}}(\xv_{(1)}^n, \cdots, \xv_{(L)}^n)$ is continuous over $\boldsymbol{\varTheta}$ and strictly positive, $\forall n \in \mathbb{N}$, $\forall (\xv_{(1)}^n, \cdots, \xv_{(L)}^n) \in \mathcal{X}^{L \times n}$.
The set $\{\Xl\}_{1\le l \le L}$ is said to be a {\em set of $L$ WP-sources} if it is a set of $L$ sources that generate random vectors according to one of the joint probability distributions $P_{\boldsymbol{\theta}}(\xv_{(1)}^n, \cdots, \xv_{(L)}^n)$.
The set $\boldsymbol{\varTheta}$ is known by the encoder and the decoder, but the current parameter $\boldsymbol{\theta}$ is unknown. 
\end{definition}

According to the above definition, the parameter vectors $\boldsymbol{\theta}$ do not depend on the sequence length $n$. As an example, Markov sources can be modeled as WP sources where $\boldsymbol{\theta}$ represents the unknown transition probabilities between symbols. 
Further, we denote $\Hs_{\boldsymbol{\theta}}\left(  \Xk | \Xjs \right)$ the conditional spectrum entropy defined in~\eqref{eq:defHs} evaluated for the conditional distribution $P_{\boldsymbol{\theta}}(\Xv_{\k}^n | \Xv_{\j}^n)$ that derives from $P_{\boldsymbol{\theta}}(\Xv_{(1)}^n, \cdots, \Xv_{(L)}^n)$.

\begin{definition}[Achievable storage and transmission rates for WP sources] \label{def:SMRA_universalcode_achievable_noniid}
Consider a set of $L$ WP sources. 
The \emph{probabilities of error} for a universal SMRA code for a WP-source are defined $\forall \boldsymbol{\theta} \in \boldsymbol{\varTheta}$ as 
\begin{align}
P_{error}^n(\boldsymbol{\theta})& = \P \left( \underset{1 \leq k \leq L}{\cup}  \left(\underset{j^* \in \Jsetk}{\cup} \left(\hat \Xv^n_{\k|\js} \neq \Xv^n_{\k} \right) \right) \right)
\end{align}
where the event probability is defined with respect to the joint probability distribution $P_{\boldsymbol{\theta}}$.

A tuple of rates $((\Sk(\boldsymbol{\theta}),( \Rkjs(\boldsymbol{\theta}) )_{j^* \in \Jsetk })_{1\le k\le L, \boldsymbol{\theta} \in \boldsymbol{\varTheta}})$ is said to be \emph{universally achievable for SMRA} if there exists a
  \emph{universal SMRA code} such that $ \forall k, j , j^*, \boldsymbol{\theta}$, 
   \begin{align}
 & \plim s_{\k}^{n}(\Xv^n_{\k}, ( \Xv^n_{\j}  )_{j }) \leq \Sk(\boldsymbol{\theta}) \\
 & \plim r_{\k|\js}^{n}(\Xv^n_{\k}, ( \Xv^n_{\j}  )_{j })\leq \Rkjs(\boldsymbol{\theta}) \\
 & \lim_{n \rightarrow \infty} P_{error}^{n}(\boldsymbol{\theta}) = 0
 \end{align}
where the event probability in the $\text{p}\text{ - }\lim\sup$ is defined with respect to the joint probability distribution $P_{\boldsymbol{\theta}}$.

\end{definition}

When the parameter vector $\boldsymbol{\theta}$ is unknown by the encoder and by the decoder, the transmission rates and the storage rate are given in the following theorem.
\begin{theorem}
Let $\{\Xl\}_{1\le l \le L}$ be a set of $L$ WP-sources (see Definition \ref{def:SPSourcesMRA}). 
Assume that the encoder can produce $\Pt$ sequences of estimated parameters $ \Theta_{\pt}^{(n)}$ such that $\forall \pt \in \{1,\cdots, \Pt\}$, $\Theta_{\pt}^{(n)}$ converges in probability to $\theta_{\pt}$.
The  tuple of rates $((\Sk(\boldsymbol{\theta}),( \Rkjs(\boldsymbol{\theta}) )_{j^* \in \Jsetk })_{1\le k\le L, \boldsymbol{\theta} \in \boldsymbol{\varTheta}})$ is achievable for SMRA (see  Definition~\ref{def:SMRA_universalcode_achievable_noniid})  if and only if $\forall k \in \llbracket 1,L \rrbracket, \forall j^{*} \in \Jsetk$, $\forall \boldsymbol{\theta} \in  \boldsymbol{\varTheta}$,
\begin{subequations}\label{eq:RS_noniid_unknown}
\begin{align}
\Rkjs(\boldsymbol{\theta}) &\ge \Hs_{\boldsymbol{\theta}} \left(  \Xk | \Xjs \right)\\[.1cm]
\Sk(\boldsymbol{\theta})&\ge \displaystyle \max_{j \in \Jsetk} \Hs_{\boldsymbol{\theta}} \left(  \Xk | \Xj \right),
\end{align}
\end{subequations}
provided  that all statistics of the sources  are known by both the encoder and the decoder, and that $\Mk$, the  size of the set of possible  previous requests $\Jsetk$ is finite.
\label{th:noniid_stat_unknown}
\end{theorem}
\begin{proof}
Throughout the proof, the notation $\Theta_n \overset{P}{\rightarrow} \theta$ denotes the convergence in probability of $\Theta_n$ to $\theta$.
For simplicity, here, we assume that $\theta_{\pt} \in [0,1]$, $\forall \pt \in \{1,\cdots, \Pt\}$, but the proof can be easily generalized to any parameter $\theta_{\pt} \geq 0$.

 {\em Learning sequence}
 For each of the $\Pt$ estimated parameters $\theta_\pt^{(n)}$ obtained from the sequences $(\mathbf{x}_1^n, \cdots, \mathbf{x}_L^n)$, the encoder constructs a learning sequence $\mathbf{s}_\pt^{u_n}$ of length $u_n$ that contains $\lfloor\theta_\pt^{(n)} u_n \rfloor$ values $1$ and $u_n-\lfloor\theta_\pt^{(n)} u_n \rfloor$ values $0$.
The parameter $\theta_\pt^{(n)}$ can hence be estimated as $\hat{\theta}^{(u_n)}_\pt = \frac{\sum_{i=1}^{u_n} s_{\pt,i}}{u_n} = \frac{\lfloor\theta_\pt^{(n)} u_n \rfloor}{u_n}$, and we denote $\SetET = [\hat{\theta}_1^{(u_n)}, \cdots, \hat{\theta}_\Pt^{(u_n)}] $.
 For source sequences of length $n$, the rate needed to store and transmit the learning sequences is given by $\frac{\Pt u_n}{n}$, since for each of the $\Pt$ parameters $\theta_\pt$, there are $2^{u_n}$ possible learning sequences. 

 \begin{lemma}\label{lem:convT}
Assume that $\underset{n \rightarrow \infty}{\lim} u_n = +\infty$. Then $\forall \pt \in \{1,\cdots, \Pt\}$, 
 $ \widehat{\Theta}^{(u_n)}_\pt \overset{P}{\rightarrow} \theta_\pt  $.
\end{lemma}

\begin{proof}
From $\hat{\theta}^{(u_n)}_\pt = \frac{\lfloor\theta_\pt^{(n)} u_n \rfloor}{u_n}$, we get the inequality $\frac{\theta_\pt^{(n)} u_n-1}{u_n} \leq \hat{\theta}^{(u_n)}_\pt \leq \frac{\theta_\pt^{(n)} u_n+1}{u_n}$, which implies
 \begin{equation}\label{eq:diffE}
  \left| \hat{\theta}^{(u_n)}_\pt - \theta_\pt^{(n)} \right| \leq \frac{1}{u_n}
 \end{equation}
 
 Let $\epsilon > 0$. Then, by marginalization,
 \begin{align}\notag
  \P(|\widehat{\Theta}^{(u_n)}_\pt - \theta_\pt |>\epsilon)  = & \P(|\widehat{\Theta}^{(u_n)}_\pt - \theta_\pt  |>\epsilon \left| ~ |\Theta_\pt^{(n)} - \theta_\pt|>\epsilon \right.) \P (|\Theta_\pt^{(n)} - \theta_\pt|>\epsilon) \\ \label{eq:dvtprobatheta}
  & + \P(|\widehat{\Theta}^{(u_n)}_\pt - \theta_\pt |>\epsilon \left| ~ |\Theta_\pt^{(n)} - \theta_\pt|\leq \epsilon \right.) \P (|\Theta_\pt^{(n)} - \theta_\pt|\leq \epsilon) .
 \end{align}
 The fact that $\Theta_\pt^{(n)}$ converges in probability to $\theta_\pt$ implies that $\forall \epsilon > 0$, $\lim_{n \rightarrow \infty} \P(|\Theta_\pt^{(n)} - \theta_\pt|>\epsilon) = 0$. 
Further, by marginalization with respect to the deterministic condition~\eqref{eq:diffE}, we get
\begin{align}\notag
 \P(|\widehat{\Theta}^{(u_n)}_\pt - \theta_\pt |>\epsilon \left| ~ |\Theta_\pt^{(n)} - \theta_\pt|\leq \epsilon \right.) & = \P(|\widehat{\Theta}^{(u_n)}_\pt - \theta_\pt |>\epsilon \left| ~ |\Theta_\pt^{(n)} - \theta_\pt |\leq \epsilon,| \widehat{\Theta}^{(u_n)}_\pt - \Theta_\pt^{(n)} | \leq \frac{1}{u_n} \right.) \\ \notag
 & = \P(|\widehat{\Theta}^{(u_n)}_\pt - \theta_\pt |>\epsilon \left| ~ | \widehat{\Theta}^{(u_n)}_\pt - \theta_\pt \right| \leq \frac{1}{u_n} + \epsilon) \\ \notag
 & = 0 .
\end{align}
As a result, from~\eqref{eq:dvtprobatheta}, $\lim_{n \rightarrow \infty} \P(|\widehat{\Theta}^{(u_n)}_\pt - \theta_\pt |>\epsilon) = 0$, which shows the convergence in probability of $\widehat{\Theta}^{(u_n)}_\pt$ to $\theta_\pt$.
\end{proof}

{\em Encoder and decoder}
We rely on the same encoders and decoders as in the proof with know statistics, except that the decoding condition is now given by
 \begin{equation}\label{eq:Adef}
  T_{\varepsilon}^{(n)}(X,X_{\j}) = \left\{ (\xv^{n},\xv_{\j}^{n}) \left| \frac{1}{n} \log \frac{1}{P_{\SetET }(\mathbf{x}^{n}|\xv_{\j}^{n})} < \sum_{i=1}^{\sigA(j)} r_i - \varepsilon \right. \right\} .
 \end{equation}
 where $P_{\SetET}(\mathbf{x}^{n}|\xv_{\j}^{n})$ is obtained by marginalization of $P_{\SetET} (\mathbf{x}_1^n, \cdots, \mathbf{x}_L^n)$. 

 From the proof for non i.i.d. sources with known statistics, we show that as $n \rightarrow \infty$,  
\begin{align}
\Rkjs &\ge \Hs_{\SetER} \left(  \Xk | \Xjs \right) + \lim_{n \rightarrow \infty} \frac{\Pt u_n}{n}  \\[.1cm]
\Sk &\ge \displaystyle \max_{j \in \Jsetk} \Hs_{\SetER} \left(  \Xk | \Xj \right) + \lim_{n \rightarrow \infty} \frac{\Pt u_n}{n} .
\end{align}
where 
\begin{equation}
  \Hs_{\SetER}\left(  \Xl | \Xj \right) = \plim \frac{1}{n} \log \frac{1}{P_{_{\SetETR}}(\Xv_{(l)}^{n}|\Xv_{(j)}^{n})} .
\end{equation}
Note that in the above expression, p  (in the $\plim$) comes from the true joint probability distribution $P_{\boldsymbol{\theta}}(\mathbf{x}_1^n, \cdots, \mathbf{x}_L^n) $, while $P_{_{\SetET}}(\xv_{(l)}^{n}|\xv_{(j)}^{n})$ comes from $P_{\SetET}(\mathbf{x}_1^n, \cdots, \mathbf{x}_L^n)$.  
 
 In order to get that $\widehat{\Theta}^{(u_n)}_\pt \overset{P}{\rightarrow} \theta_\pt$ from Lemma~\ref{lem:convT}, we choose the sequence $u_n$ such that $\lim_{n \rightarrow \infty} u_n = +\infty$. 
In addition, for the excess rate $ \frac{\Pt u_n}{n}$ to vanish as $n$ goes to infinity, we set $u_n = o(n)$. 
Thus, in order to prove that unknown statistics do not induce any rate loss, we want to show that
\begin{equation}
 \Hs_{\SetER}\left(  \Xl | \Xj \right) =  \Hs_{\boldsymbol{\theta}}\left(  \Xl | \Xj \right), 
\end{equation}

In the following, we remove the index $\ell$ for simplicity.
Let 
\begin{align}
\An & = \frac{1}{n} \log \frac{1}{P_{\SetETR}(\mathbf{X}^{n}|\Xv_{\j}^{n})} \\
 \Bn & = \frac{1}{n} \log \frac{1}{P(\mathbf{X}^{n}|\Xv_{\j}^{n})}  \\
 \Vn & = \An - \Bn
\end{align}
\begin{lemma}\label{lem:Vnconv}
$\forall \epsilon> 0$, $\lim_{n \rightarrow \infty} \P(|\Vn|>\epsilon) = 0$ ($\Vn$ converges in probability to $0$).  
\end{lemma}
\begin{proof} The function  $\mathbf{t} \mapsto P_{\mathbf{t}}(\mathbf{x}_1^n, \cdots, \mathbf{x}_L^n)$ is continuous, which implies that the function $\mathbf{t} \mapsto P_{\mathbf{t}}(\mathbf{x}|\xv_{\j}^{n})$ is also continuous, since $P_{\mathbf{t}}(\xv_{\j}^{n})>0$ (by strict positivity of $f_{\mathbf{t}}(\mathbf{x}_1^n, \cdots, \mathbf{x}_L^n)$). The function $y\rightarrow \frac{1}{n} \log \frac{1}{y}$ is also continuous.
Let $\epsilon_2>0$. Then by continuity  $\forall n \in \mathbb{N}$, $\forall \xv^{n}, \xv_{\j}^{n} \in \mathcal{X}^n$ of $\mathbf{t} \rightarrow \frac{1}{n} \log \frac{1}{P_{\mathbf{t}}(\xv^{n} | \xv_{\j}^{n})} $ ,  $\exists \epsilon_1>0$ such that 
\begin{equation}\label{eq:continuityT}
 \left\| \boldsymbol{\theta}_1 - \boldsymbol{\theta}_2 \right\| < \epsilon_1 \Rightarrow \forall n\in \mathbb{N}, \forall \xv^{n}, \xv_{\j}^{n} \in \mathcal{X}^n, \left| \frac{1}{n} \log \frac{1}{P_{\boldsymbol{\theta}_1}(\xv^{n} | \xv_{\j}^{n})} -  \frac{1}{n} \log \frac{1}{P_{\boldsymbol{\theta}_2}(\xv^{n} | \xv_{\j}^{n})} \right|  \leq \epsilon_2
\end{equation}
Then, by marginalization, \small
\begin{align}\notag
 \P(|\Vn|>\epsilon_2) = & \P(|\Vn|>\epsilon_2 \left| ~ \| \SetETR - \boldsymbol{\theta} \| \geq \epsilon_1 \right.) \P(\| \SetETR - \boldsymbol{\theta} \| \geq \epsilon_1) \\
 & + \P(|\Vn|>\epsilon_2 \left| ~ \| \SetETR - \boldsymbol{\theta} \| < \epsilon_1 \right. ) \P(\| \SetETR - \boldsymbol{\theta} \| < \epsilon_1)
\end{align}\normalsize
From condition~\eqref{eq:continuityT}, $ \P(|\Vn|>\epsilon_2 \left| ~ \| \SetETR - \boldsymbol{\theta} \| < \epsilon_1 \right. ) = 0 $. From Lemma~\ref{lem:convT}, $\lim_{n\rightarrow \infty}  \P(\| \SetETR - \boldsymbol{\theta} \| \geq \epsilon_1) = 0$.
As a result, $\lim_{n\rightarrow \infty} \P(|\Vn|>\epsilon_2) = 0 $, which completes the proof of the lemma. 
\end{proof}

\begin{lemma}\label{lem:AnBn}
 $\plim B_n = \plim A_n$ .
\end{lemma}
\begin{proof}
 Let $\bar{\beta}$ be the $\inf$ of values $\beta$ such that $\lim_{n\rightarrow \infty} \P(\Bn>\beta)=0$ and let $\bar{\alpha}$ be the $\inf$ of values $\alpha$ such that $\lim_{n\rightarrow \infty} \P(\An)>\alpha)=0$.
 
 Let $\epsilon>0$. By marginalization, \small
 \begin{align}\notag
  \P(\An > \bar{\beta} + 2\epsilon) = & \P(\An > \bar{\beta} + 2\epsilon \left| ~ |\Vn| > \epsilon\right.) P (|\Vn| > \epsilon ) \\ 
 +    &    \P(\An > \bar{\beta} + 2\epsilon \left| ~ |\Vn| \leq \epsilon \right.) \P (|\Vn| \leq \epsilon)
 \end{align}\normalsize
From Lemma~\ref{lem:Vnconv}, $\lim_{n \rightarrow \infty} \P(|\Vn|>\epsilon) = 0$. In addition,
\begin{equation}
\P(\An > \bar{\beta} + 2\epsilon \left| ~ |\Vn| \leq \epsilon \right.) \leq \P(\Bn > \bar{\beta})
\end{equation}
Since $\lim_{n\rightarrow \infty} \P(\Bn>\beta)=0$, we get $\lim_{n\rightarrow \infty} \P(\An > \bar{\beta} + 2\epsilon)=0$, which shows that $ \bar{\alpha} \leq \bar{\beta} + 2\epsilon$. 
 
 By showing in the same manner that $\lim_{n \rightarrow \infty} \P\left(\An > \bar{\beta} - 2\epsilon\right) > 0$, we get that $ \bar{\beta} - 2\epsilon \leq \bar{\alpha}$. 
 As a result, $\bar{\beta} - 2\epsilon \leq \bar{\alpha} \leq \bar{\beta} + 2\epsilon$. Since this inequality is true $\forall \epsilon>0$, we can conclude that $\bar{\beta} = \bar{\alpha} $. 
\end{proof}
From Lemma~\ref{lem:AnBn}, we can deduce that $\Hs_{\SetER}\left(  \Xl | \Xj \right) =  \Hs_{\boldsymbol{\theta}}\left(  \Xl | \Xj \right)$, which concludes the proof.
 \end{proof}
 The results of Theorem~\ref{th:noniid_stat_unknown} show that when the parameter vector $\boldsymbol{\theta}$ is unknown, the storage and transmission rates are the same as if the parameters were known. 
	
%
\section{Optimal transmission and storage Rates-Distortion regions}\label{sec:lossy}
In this section, we describe lossy source coding for SMRA.
We now allow some distortion $d(\Xk,\hat{X}_{(k)}) $ between the requested source $\Xk$ and its reconstruction $\hat{X}_{(k)}$.
The aim is to determine the minimum storage rate $\Sk(\delta)$ and transmission rate $\Rkjs(\delta)$ that can be achieved with respect to a distortion constraint $\delta$.
This problem is addressed in the simpler case of i.i.d. Gaussian sources.

\begin{definition}[i.i.d. Gaussian sources]~\label{def:iidGaussianMRA}
The set $\{\Xl\}_{1\le \ell \le L}$ is said to be a {\em set of $L$ i.i.d. Gaussian sources} if the joint distribution can be factorized as:
\begin{equation}
P(\xv^n_{(1)},...,\xv^n_{\l},...,\xv^n_{\L}) = \prod_{i=1}^n P(x_{(1),i},...,x_{\L,i})
\end{equation}
and if each pair of symbols $(X_{(k),i},X_{(j),i})$ is jointly Gaussian with covariance matrix
\begin{equation}
 \Ckj = \begin{bmatrix}
      \sigx^2 & \cc \sigx \sigy \\
      \cc \sigx \sigy & \sigy^2
     \end{bmatrix}
\end{equation}
\end{definition}
Since the sources are Gaussian, we consider a quadratic distortion measure $d(\Xk,\hat{X}_{(k)})  = (\Xk-\hat{X}_{(k)})^2$.

As another difficulty of the lossy source coding problem, the source $\Xj$ that serves as side information for $\Xk$ was also reconstructed with a certain distortion.
As a first attempt to address this problem, we reinterpret the SMRA framework as follows. 
For source $\Xk$, we consider an extended set $\Jsetkd$ that includes all the possible distorded versions of the sources that can be available at the decoder when $\Xk$ is requested. 
This increases the size of $\Jsetkd$ compared to $\Jsetk$ since the distortion in the previous request depends on the path that was followed by the user in the navigation graph.
As a result, each of the possible previous requests given by the navigation graph will appear several times in $\Jsetkd$, with different distortion levels.

Although the set $\Jsetkd$ may be difficult to characterize in practical situations, we consider it here in order to provide insights on the tradeoff between transmission-storage rate versus distortion in the lossy context. 
In addition, the Gaussian model of Definition~\ref{def:iidGaussianMRA} is hereafter assumed to be between $\Xk$ and the distorded sources contained in $\Jsetkd$. 
As a result, the parameters $\sigma_j^2$ and $\rho_{k,j}$ will depend on the distortion levels in the $\Xj \in \Jsetkd$.
Although not completely rigorous, we consider this assumption here in order to obtain a a first characterization of the rate-storage tradeoff in SMRA.  

Under the above assumptions, the transmission-storage rates versus distortion regions for Gaussian sources for SMRA are given in Theorem~\ref{th:rd_stat_known}.
\begin{theorem}
Let $\{\Xl\}_{1\le l \le L}$ be a set of $L$ i.i.d. Gaussian sources (see Definition~\ref{def:iidGaussianMRA}).
For a given parameter $\delta$, 
the  tuple of rates $((\Sk(\delta),( \Rkjs(\delta) )_{j^* \in \Jsetkd })_{1\le k\le L})$ is achievable for SMRA if and only if $\forall k \in \llbracket 1,L \rrbracket, \forall j^{*} \in \Jsetkd$
\begin{align}
\begin{array}{ll}
\Rkjs(\delta) &\geq R_{k,j^{*}}(D_{k,j^{*}})   \\
\Sk(\delta) &\geq \displaystyle \max_{j \in \Jsetkd} R_{k,j}(D_{k,j}) \\ 
 E[(\hat{X}_{(k)} - \Xk)^2|\Xjs] & \leq D_{k,j^{*}}
\end{array}
\label{eq:RSWZ}
\end{align}
where
\begin{equation}
 R_{k,j}(D_{k,j})= \left\{ \begin{array}{ll}
                               \frac{1}{2} \log_2 \frac{\sigkj^2}{D_{k,j}} & \text{ if } \sigkj^2 > D_{k,j} ,\\
                               0 & \text{ otherwise,}                
 \end{array}
 \right. 
\end{equation}
with $D_{k,j} = \frac{\delta \sigkj^2}{\sigkj^2 + \delta\cc^2}$ and $\sigkj^2 = \sigx^2 (1-\cc^2)$,
provided  that all statistics of the sources  are known by both the encoder and the decoder, and that $\Mk$, the  size of the set $\Jsetkd$ is finite.
\label{th:rd_stat_known}
\end{theorem}

\textit{Remark:}
The parameter $\delta$ is given by $\delta = E[(\hat{X}_{(k)} - \Xk)^2] $ when no previous request is available at the decoder. 
When a previous source $\Xj$ is available (in principle a distorded version of a possible previous request), the source $\Xk$ can be reconstructed with a distortion $D_{k,j} \leq \delta$ that depends on the parameter $\delta$ and on the source statistics $\cc$, $\sigx $, $\sigy$.
The parameter $\delta$ hence leads to different distortions $D_{k,j}$, depending on the available source $\Xj $.
This fact was also observed in~\cite{tian2007multistage,watanabe14IT} where side informations with different statistics are available at different decoders.

\textit{Remark:}
In~\eqref{eq:RSWZ}, the transmission rate $\Rkjs(\delta) = \frac{1}{2} \log_2 \frac{\sigkjs^2}{D_{k,j^{*}}}$ corresponds to the Wyner-Ziv rate-distortion function for a given target distortion $D_{k,j^{*}}$ when $\Xjs$ is the only possible side information.
The storage rate $\Sk(\delta)$ is still given by the worst possible rate-distortion function $\max_{j \in \Jsetkd} R_{k,j}(D_{k,j})$. 
The tradeoff between the transmission-storage rates and the distortions is adressed through the parameter $\delta$ that affects all the distortion levels $D_{k,j}$. 
%
 \begin{proof}
 {\em Source ordering.} For each source index $k$, let 
\begin{subequations}\label{eq:orderingMaprd}
\begin{align}
\sigA: \Jsetkd &\to \llbracket 1, \Mk \rrbracket \\
j & \mapsto \sigA(j)
\end{align}
\end{subequations}
be an ordering of the source indexes in $\Jsetkd$ such that $m=\sigA(j)$ and
\begin{align}
&\!\!\!\!\!  \rho_{k,\sigA^{-1}(1)} \leq \cdots \leq \rho_{k,\sigA^{-1}(m)} \leq \cdots \leq \rho_{k,\sigA^{-1}(|\Jsetkd|)} .
\label{eq:orderSourcePreviousRequestRD}
\end{align}
 
  We keep the same notations and conventions as for the proof of Theorem~\ref{th:iid_stat_known}.
 In particular, we omit the index $(k)$ of the source to be encoded.
 The proof involves two proof techniques: Gaussian test channels and typical sequences.
 This enables to derive the rate-distortion functions in the simpler Gaussian case, but also to deal with incremental coding that is used to transmit the codeword from the encoder to the decoder.
 
  \textbf{Achievability :}
{\em Random code generation.} Let $U$ be a random variable such that $U = \gamma X + \Phi$, where $\Phi \sim \mathcal{N}(0,\gamma \delta)$ is independent of $X$ and $\gamma = 1 - \frac{\delta}{\sigx^2}$.
This test channel comes from~\cite{Draper04}.
Generate $2^{nr_0}$ sequences $\Un$ at random, where $r_0 = I(X;U) + \epsilon$.
Denote by $\mathcal{C}$ the set of generated sequences $\un(s)$ and index them with $s \in \lb 1,2^{nr_{0}}\rb$. 
Assign each $\un \in \mathcal{C}$ to incremental bins, following the same process as in the proof of Theorem~\ref{th:iid_stat_known}.
The size of the $\M$ successive bins depends on the values $r_{i}$ such that $r_1 + \cdots + r_{\sigA(j)} = I(X;U) - I(\Xj;U) + 5\epsilon$.
It can be verified that $\forall j$, $r_{\sigA(j)}\geq0$ (condition on successive $\cc$ of Theorem~\ref{th:rd_stat_known}).
This defines mappings $f_{\sigA(j)}(\un) = (i_1,\cdots,i_{\sigA(j)})$ where the $i_{\sigA(j)}$ are the indices of the successive bins to which $\un$ belongs.

{\em Encoding.} 
Given a sequence $\xv^n$, find a sequence $\un(s) \in \mathcal{C}$ such that $(\xv^n, \un(s)) \in A_{\epsilon}^{(n)} (X, U)$. 
The offline encoder then sends to the storage unit the index sequence $i_1,\cdots,i_{\M} $ to which $\un(s)$ belongs.
Upon request of the source $X$ and previous request $j$, the online extractor sends to the user the index sequence $ (i_1,\cdots,i_{\sigA(j)})$ to which $\un(s)$ belongs.

{\em Decoding. }
Given the received index sequence $(i_{1},...,i_{\sigA(j)})$ and the side information $\xv^n_{\j}$, declare $\hat \un = \un(s)$ if there is a unique pair of sequences $(\un(s), \xv^n_{\j})$ such that $f_{{\sigA\j}}(\un(s)) =(i_{1},...,i_{\sigA\j})$ and  $(\un(s), \xv^n_{\j}) \in A^n_\epsilon(U, X_{\j})$, where $A^n_\epsilon$ is defined in~\eqref{eq:jointtyp}.
Then compute $\hat{X} =  \alpha U + \beta \Xj$, with $\alpha = \frac{\sigkj^2}{\gamma\sigkj^2 + \delta  }$ and $\beta = \frac{\cc \sigx}{\sigy} \frac{\delta}{\gamma\sigkj^2 + \delta  }$.

{\em Probability of error. }
We define the error events: 
\begin{align}
E_{0,0}&= \{ (\Xv^n,\Xv^n_{\j}) \notin A^n_\epsilon(X, X_{\j})\} \\
E_{0,1}&= \{ (\Xv^n,\Xv^n_{\j}) \in A^n_\epsilon(X, X_{\j})  \nonumber \\ 
& \ \ \ \ \ \ \ \ \ \ \ \  \mbox{ but } \nexists s \mbox{ such that } (\Xv^n,\un(s)) \in A^n_\epsilon(X, U)\} \\
E_{0,2}&= \{ (\Xv^n,\un(s)) \in A^n_\epsilon(X, U) \nonumber \\
& \ \ \ \ \ \ \ \ \ \ \ \  \mbox{ but } \nexists s \mbox{ such that } (\Xv^n_{\j},\un(s)) \in A^n_\epsilon(X, U)\} \\
E_j&= \{ \exists s' \neq s: f_{\sigA\j} (\un(s')) =  f_{\sigA\j} (\un(s))   \nonumber \\
& \ \ \ \ \ \ \ \ \ \ \ \  \mbox{ and } (\Xv^n_{\j},\un(s')) \in A^n_\epsilon(X_{\j},U)\}, \ \ \ \forall j \in \Jc
\end{align}

First, using the same arguments as in the proof of Theorem~\ref{th:iid_stat_known}, we have $\P ( E_{0,0} ) \to 0$ as $n\to\infty$.
Second, $\P ( E_{0,1} ) \to 0$ as $n\to\infty$ by the source coding theorem and since $r_0 > I(U;X)$.
Then, $\P ( E_{0,2} ) \to 0$ as $n\to\infty$ by the Markov Lemma.
We now evaluate $E_j$, $\forall j \in \Jc$
\begin{align}
\P ( E_j )&  = \sum_{{(\un,\xv^n_{\j})  \in  \mathcal{C}\times \Xc^n_{\j} } } P ( \un,\xv^n_{\j} ) ...&\\
& \P 
\Big( \exists s' \neq s: f_{\sigA\j} (\un(s')) =  f_{\sigA\j} (\un(s))  \mbox{ and } (\un(s'),\xv^n_{\j}) \in A^n_\epsilon
\Big) \nonumber \\
&  \le \sum_{(\un,\xv^n_{\j})} P ( \un,\xv^n_{\j} ) \sum_{\mathclap{\substack{s' \neq s \\ f_{\sigA\j} (\un(s')) =  f_{\sigA\j} (\un(s))}}}
\P \Big(  (\un(s'),\xv^n_{\j}) \in A^n_\epsilon   \Big) \label{eq:EjWZ_1}\\
&\le 2^{ -n \big( r_{0}- (r_1 +  ... +r_{\sigA\j})\big)}     .2^{n\Large(I(U;X_{\j})) - 3\epsilon \Large) }\label{eq:EjWZ_3}
\end{align}
Hence $\forall j \in \Jc$, $\P ( E_{j} ) \to 0$ as $n\to\infty$ since we have $r_0 - (r_1 +  ... +r_{\sigA\j}) < I(U;X_{j}) - 3\epsilon$.
At the end, $P_{error}^n = \P ( E_{0,0} \cup E_{0,1} \cup E_{0,2} \cup  \bigcup_{j\in \Jc}  E_j)\to 0$ as $n\to\infty$.

{\em Distortion. }
With the above parameters $\alpha$, $\beta$, and $\delta$, the actual distortion can be calculated as
\begin{align}
 D_{j} & =  E[(\hat{X}- X)^2 | \Xj] \notag \\
 & = (1- P_{error}^n)\frac{\delta \sigkj^2}{\sigkj^2 + d\cc^2} + P_{error}^n \delta_{max}
\end{align}
where $\delta_{max}$ is a constant that represents the maximum possible distortion.
As a result, $ D_{j} \to \frac{\delta \sigkj^2}{\sigkj^2 + \delta\cc^2}$ as $n\to\infty$.

In other words, for any previously requested source $j \in \Jc$, there exists a code with a distortion $D_{j} = \frac{\delta \sigkj^2}{\sigkj^2 + \delta\cc^2}$ and compression rate $r_{1}+  ... +r_{\sigA\j}$, provided that $r_{1}+  ... +r_{\sigA\j} \ge I(X;U) - I(U;\Xj) = \frac{1}{2}\log_2 \frac{\sigkj^2}{D_{j}}  $ and that $M$ is finite. 
Finally, to be able to serve any request $j$ with distortion $D_{j}$, the compression scheme needs to prepare a code for any $j \in \Jc$. Therefore, the storage requires  $S \ge  \max_{j \in \Jc} \frac{1}{2}\log_2 \frac{\sigkj^2}{D_{j}}$.
  Moreover, if the actual previous request is $j^*$, to achieve the distortion $D_{j^{*}}$, the transmission rate $R$ needs to satisfy $R  \ge-\frac{1}{2}\log_2 \frac{\sigkjs^2}{D_{j^{*}}}$,  which completes the achievability proof. 

\textbf{Converse}: The converse can be done by contradiction as in the lossless cases.
 \end{proof}
 
%

 \textit{Remark:} Theorem~\ref{th:rd_stat_known} still holds when the source statistics (the variances and the correlation coefficient) are unknown.
 The lack of knowledge on the joint statistics between $\un$ and $\xv^n_{\j}$ can be handled by the method of types, as described in the proof of Theorem~\ref{th:iid_stat_unknown}.
 The parameters $\sigx$ and $\sigy$ can be estimated online at the encoder and at the decoder, respectively. The parameter $\cc$ can be estimated from the relation $E[uy] = \gamma \cc \sigx \sigy $.

%
\section{Generalization to unbounded memory at the decoder}
\label{sec:RSboundsMRA:iid}

In Sections~\ref{sec:RSbounds:iid} to \ref{sec:lossy}, one considers that only the previously requested source is stored in the decoder memory.  In this section, Theorem~\ref{lem:manySourcesAtDecoder} provides the gain that can be obtained when all previously requested sources are stored at the decoder. 

\begin{definition}[Set of possible previous requests] Let $k$ denote the current request.
The \emph{set of possible previous requests} $\Ffamk=\{\Ec_{1},\ldots, \Ec_{\Mk} \} $  is {a set of subsets} of $\lb 0, L \rb /\{k\}$, where each subset contains the previous requests that might be available at the decoder. $\Mk=|\Ffamk|$. 
\label{def:FamilyPreviousRequests}
\end{definition}

\begin{figure}[h]\begin{center}
\includegraphics[width=0.3\linewidth]{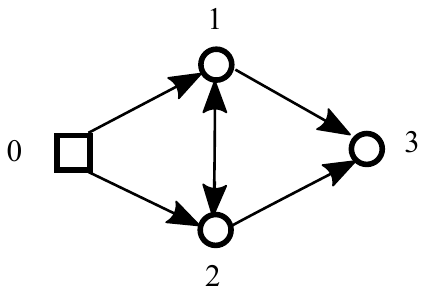} 
\end{center}
\caption{Graph example, where $0$ stands for the initial node. $\Jc_{(1)}=\{0,2\}$ and $\Jc_{(3)}=\{1,2\}$.}
\label{fig:ExGraph}
\end{figure}

\textit{Example:}
To illustrate $\Ffamk$ on a simple example, consider the directed graph in Figure~\ref{fig:ExGraph}. Here, $\Fc_{(2)}=\{\{0\}, \{0,1\}\}$, and  $\Fc_{(3)}=\{ \{0,1\},\{0,2\}, \{0,1,2\}\}$.
Note that the graph is directed giving rise to ordered sequences, but the subsets are specified without any order, as the order does not matter in the conditional entropy. Therefore different paths in the directed graph may lead to the same subset in $\Ffamk$. 

\begin{theorem}[SMRA while storing all the previously requested sources at the decoder]
Let $\{\Xl\}_{1\le l \le L}$ be a set of $L$ discrete  i.i.d. sources.
The  tuple of rates $((\Sk,( \Rkjs )_{j^* \in \Jsetk })_{1\le k\le L})$ is achievable for SMRA  if and only if $\forall k \in \llbracket 1,L \rrbracket, \forall \Ec^{*} \in \Ffamk$
\begin{subequations}
\begin{align}
\RkEjs &\ge H \left(  \Xk | X_{\Ejs} \right),   \\
\Sk&\ge \displaystyle \max_{\Ec \in \Ffamk} H \left(  \Xk | X_{\Ec} \right).
\end{align}
\label{eq:RSmany}
\end{subequations}
provided that all statistics of the sources  are known by both the encoder and the decoder, that all the previously requested sources are stored at the decoder, and that $|\Ffamk| \} $ is finite.\\
\label{lem:manySourcesAtDecoder}
\end{theorem}

\textit{Remark:} Theorem~\ref{lem:manySourcesAtDecoder} allows one to compare the case where only one single source $j^*$ is known at the decoder  (Theorem~\ref{th:iid_stat_known}), with the more general case of encoding along an arbitrary path of the navigation graph, where many sources $\Ejs$ are available at the decoder (Theorem~\ref{lem:manySourcesAtDecoder}). 
As intuition suggests, taking into account more sources at the decoder, allows to reduce the rate at which $\Xk$ is stored and transmitted  since:
\begin{subequations}
\begin{align}
 H \left(  \Xk | X_{\Ejs} \right) & \le H \left(  \Xk | X_{(j^*)} \right) ,   \\
\max_{\Ec \in \Ffamk} H \left(  \Xk | X_{\Ec} \right) & \le \max_{j \in \Jsetk} H \left(  \Xk | \Xj \right).
\end{align}
\label{eq:RS1vsmany}
\end{subequations}
where $j^* \in \Ejs$ is the request preceding $k$ and $\Jsetk$ (defined in Section \ref{sec:MRAdefinition}) is the one-hop neighborhood of the source with index $k$.
Equality in (\ref{eq:RS1vsmany}a)   and (\ref{eq:RS1vsmany}b) occurs if all  
the nodes of the one-hop neighborhood $\Jsetk$ are all connected to the initial node $0$.

\begin{proof}
Let $Y$ be a sufficient statistic of $X_{\Ejs}$ for $\Xk$, such that 
\begin{align}
H \left(  \Xk | Y \right)=H \left(  \Xk | X_{\Ejs} \right).
\label{eq:condEneq}
\end{align}
Let us apply Theorem~\ref{th:iid_stat_known} with $Y$ as the previous source available at the decoder.
Then, applying the conditional entropy equality \eqref{eq:condEneq}  leads to \eqref{eq:RSmany}.

Note that, by applying Theorem~7,  we also construct, for each source of index $k$, a reordering of the  family of subsets of previous requests (Definition~\ref{def:FamilyPreviousRequests})  according to
\begin{align}
&\!\!\!\!\!  H (  \Xk | X_{\Ec_{1}} ) \le ...  \le H (  \Xk | X_{\Ec_{|\Ffamk|}} ).
\label{eq:orderSourcePreviousRequestSet}
\end{align}

We now show (\ref{eq:RS1vsmany}). By construction, the previously requested source $X_{(j^*)}$ belongs to the set of previously requested sources $X_{\Ejs}$. So, since conditioning reduces entropy, we have (\ref{eq:RS1vsmany}a).

Moreover, we have 
\begin{align}
 \forall \Ec \in \Ffamk, \forall l \in \Ec, H \left(  \Xk | X_{\Ec} \right) &\le  H \left(  \Xk | \Xl \right), 
 \label{eq:HsetH1}
\end{align}
which holds in particular for any source that belongs to $\Ec$ and to the one-hop neighborhood of $\Xk$, {\it i.e.} it holds for any source indexed by $l\in \Ec\cap\Jsetk$. Note that the set $\Ec\cap\Jsetk$ may contain many indexes since each set $\Ec$ corresponds to eventually many paths, as a path is ordered whereas the set is not. Finally, we get
\begin{align}
\max_{\Ec \in \Ffamk} H \left(  \Xk | X_{\Ec} \right) &\le \max_{\Ec \in \Ffamk} \max_{l \in \Ec\cap\Jsetk} H \left(  \Xk | \Xl \right) \label{eq:upperboundHa}\\
&= \max_{l \in \Jsetk} H \left(  \Xk | \Xl \right) 
\label{eq:upperboundH}
\end{align}
where the second equality follows from the fact that $\cup_{\Ec \in \Ffamk} (\Ec\cap \Jsetk) = \Jsetk$ {\it i.e.} the multi-hop neighborhood of a node $k$ reduced to one-hop is exactly the  
one-hop neighborhood $\Jsetk$. This shows (\ref{eq:RS1vsmany}b).

{\em Case of equality in (\ref{eq:RS1vsmany}b)}. If the nodes of the one-hop neighborhood $\Jsetk$ are all connected to the initial node $0$ (meaning that the navigation can start at any source 
indexed by  $\Jsetk$), then the equality holds in \eqref{eq:upperboundHa}  and therefore in  (\ref{eq:RS1vsmany}b). The equality is also achieved if the node in the one-hop neighborhood $\Jsetk$ that achieves the maximal entropy in \eqref{eq:upperboundH} is connected to node $0$.
\end{proof}

%

\section{Practical scheme}
\label{sec:channelcoding}


The encoding and decoding processes detailed in Section~\ref{sec:RSbounds:iid} cannot be used in practice as they  rely on random code generation and typical decoding. However, we can construct practical codes that approach the rate-storage bounds in the case  of discrete i.i.d. sources, when the correlation between the sources is modeled as an additive channel. This is a consequence of the following facts:
\begin{itemize}
\item[(i)]  \emph{Information theoretical duality between common broadcast channel and universal source coding with side information at the receiver}: \cite{Draper07} shows that a code optimal for common broadcasting \cite{Feder02ITW} \cite[Chap. 4]{Shulman03phd} (a common information is to be sent to a set of receivers with different channel capacities and a receiver can quit  when it receives the information needed to reconstruct the data) can be turned into an optimal code for source coding with side information and feedback link to stop the transmission.
\item[(ii)] \emph{Information theoretical duality between universal source coding with side information at the receiver and SMRA}: Section~\ref{sec:RSbounds:iid} shows that this same code \cite{Feder02ITW} \cite[Chap. 4]{Shulman03phd} can be used for SMRA, but with a novel way to determine the transmitted rate, computed at the encoder directly and based on the source ordering defined in the proof of Theorem~\ref{th:iid_stat_known}.
\item[(iii)] \emph{Existence of practical codes for the common broadcast channel}: practical channel coding schemes have been proposed for the common broadcast channel and are usually referred to as Fountain codes \cite{MacKay03ITbook,MacKay05Fountain} 
or rateless codes. These channel codes have the ability to compute repair symbols on the fly in order to meet the channel capacity constraint of each receiver.
\item[(iv)]  \emph{Channel codes can be turned into Slepian-Wolf source codes \cite{Slepian73IT}}: one can use the parity bits of a channel block code to perform a lossless compression with side information at the decoder \cite{Wyner74IT} \cite[Chap.~6]{DragottiBook09DSC}. 
\end{itemize}

We now detail this approach in the context of SMRA with discrete i.i.d. sources. First, the correlations between the source to be encoded $\Xk$ and the side informations $\Xj, j\in \Jsetk$ is modeled as a set of channels with the same input $\Xk$ and output $\Xj$
but with different statistics. Then,  an incremental and systematic channel code is constructed for this set of channels with rates $\{R_{\pi(j)}^{ch}\}_{j\in \Jsetk}$, where $R_{\pi(j)}^{ch}$ stands for the ratio between the number of useful bits $n$ and the codeword length say $n_{\pi(j)}$ for the ${\pi(j)}^{th}$ channel. This code maps the sequence $\xv^n_{\k}$ to an incremental codeword having $\xv^n_{\k}$ as systematic symbols and  $\pv_{\k}$ as parity symbols. The parity vectors are sequential:  $\pv_{\k}=(\pv^{nr_1}_{\k}, \pv^{nr_2}_{\k},... , \pv^{nr_{M_{\k}}}_{\k})$, where the sum rate is given by $\sum_{i=1}^{M_{\k}} r_i \ge\Sk$ defined in \eqref{eq:RS}. The compression rates $r_m$ and channel codes $R_m^{ch}$ are related through
\begin{align}
\sum_{i=1}^{\pi(j)} r_i & = \frac{n_{\pi(j)}-n}{n} =\frac{1}{R_{\pi(j)}^{ch}} -1 , \forall j \in \Jsetk.
\end{align}
The whole parity vector is stored with markers to identify the end of each subvector   $ \pv^{nr_{\pi(j)}}_{\k}$. Then, upon request of source  $\k$ while the previous request was $\js$, the transmitter sends the vector $(\pv^{nr_1}_{\k}, \pv^{nr_2}_{\k},... , \pv^{nr_{\pi(j^*)}})$.
Finally, the storage and transmission cost can be related to the channel rates:
\begin{align}
\mbox{practical transmission rate}&=\sum_{i=1}^{\pi(j^*)} r_i =\frac{1}{R_{\pi(j^*)}^{ch}} -1\\
\mbox{practical storage rate}&=\sum_{i=1}^{M_{\k}} r_i =\frac{1}{R_{M_{\k}}^{ch}} -1
\end{align}
Following the proof in \cite{Wyner74IT}, one can show that if the channel code achieves the capacity of the equivalent channel, then the practical scheme will also achieve the bounds derived in Theorem~\ref{th:iid_stat_known}.
    In practice, the sequential parity vector can be obtained from LDPC code constructions such as LDPC staircase codes and LDPCA codes, as we illustrate in the next section.

{The coding strategy relies on the fact that the probability of decoding error converges to $0$ when $n \rightarrow \infty$. In other words, the probability of error may not always be $0$ when encoding a finite length source, which could be damageable when using, in turn, this source as side information for the decoding of another one (consecutively requested by the same client). To circumvent this issue in practice, we propose to simulate the decoding at the encoding stage and to add extra parity bits until the probability of error actually reaches 0. This results in extra storage and transmission rates but prevents any error propagation during the client's navigation. However, due to the optimal performances obtained in the theoretical study (see Theorem \ref{th:iid_stat_known}), we know that this extra rate converges to $0$ when $n \rightarrow \infty$.}

\begin{table*}[h]
\begin{center}
  \begin{tabular}{*{7}{c|}{c}}
\hline
 & \multicolumn{6}{c|}{Number of  symbols sent}& Number\\ 
 & $p_{j_1} = 0.01$ & $p_{j_2} = 0.05$ & $p_{j_3} = 0.1$  &$p_{j_4} = 0.15$  & $p_{j_5} = 0.2$ & $p_{j_6} = 0.25$  & Stored \\
\hline
theoretical & 400 & 2,000 & 4,000& 6,000 & 8,000 & 10,000 & 40,000 \\
\hline
LDPC-& 423 & 2,023 & 4,023 & 6,023 & 8,023 & 10,023 & 40,023\\
Staircase &&&&&&& 
\vspace{0.1cm}
  \end{tabular}
        \centerline{$n=40000$}
      \caption{Comparison between theoretical and practical rate-storage performance for the coding of a source $\Xk$  made of $n$ symbols. The set of possible side information sources $\Jsetk$ is made of $6$ sources, with different erasure probability. The results were obtained with a LDPC staircase coder \cite{Roca12RFC6816}.}
  \label{table:erasure}
  \end{center}
\end{table*}
\begin{table*}[h]
\begin{center}
  \begin{tabular}{*{7}{c|}{c}}
\hline
 & \multicolumn{6}{c|}{Number of  symbols sent}& Number\\ 
 & $p_{j_1} = 0.01$ & $p_{j_2} = 0.05$ & $p_{j_3} = 0.1$  &$p_{j_4} = 0.15$  & $p_{j_5} = 0.2$ & $p_{j_6} = 0.25$  &Stored \\
\hline
theoretical & 32  &  114 & 186  & 242 & 286  & 322  & 322  \\
\hline
LDPCA    & 52  & 147  & 217  & 283 & 332  &  377 & 377 \\
  &   &   &   & &   &   &   
\vspace{0.1cm}
  \end{tabular}
      \centerline{(a) $n=396$}
      \vspace{0.1cm}

  \begin{tabular}{*{7}{c|}{c}}
\hline
 & \multicolumn{6}{c|}{Number of  symbols sent}& Number\\ 
 & $p_{j_1} = 0.01$ & $p_{j_2} = 0.05$ & $p_{j_3} = 0.1$  &$p_{j_4} = 0.15$  & $p_{j_5} = 0.2$ & $p_{j_6} = 0.25$ &Stored \\
\hline
theoretical & 512  &  1815 & 2972  & 3864  & 4575  & 5141  & 5141  \\
\hline
LDPCA    & 913  &  2160 & 3325  & 4264 & 5194  & 6329  & 6329 \\
 &   &   &   & &   &   &   
\vspace{0.1cm}
  \end{tabular}
        \centerline{(b) $n=6336$}
      \caption{Comparison between theoretical and practical rate-storage performance for the coding of a source $\Xk$  made of $n$ symbols. The set of possible side information sources $\Jsetk$ is made of $6$ sources, with different error probability. The results were obtained with a LDPCA coder \cite{Varodayan06Eurasip}.}
  \label{tab:bsc}
  \end{center}
\end{table*}

\section{Experimental validation}
\label{sec:exp}
\subsection{Experimental setup}
The rate-storage bounds derived in this paper state the performance that a SMRA scheme can theoretically achieve. 
Building SMRA codes for general sources is another complex research problem and we leave it for future works.
However, in this section, we show that for binary i.i.d. uniform sources,  already existing codes, namely LDPC staircase \cite{Roca12RFC6816} and LDPCA \cite{Varodayan06Eurasip}, can help building efficient practical implementation of SMRA schemes. 
For this experimental validation, we thus assume that the $\Xl$ are binary i.i.d. uniform sources. As in the theoretical results derived in previous section, let us focus on the performance of a single source $\Xk$ that is assumed to be requested by the user. 
For binary i.i.d. sources, we consider two forms for the conditional probability distribution $P(x_{(k)} | x_{(j)})$ (with $j \in \Jsetk$) that correspond to either erasures or binary errors.
In the next two sections, we study each of the two distributions and propose a practical SMRA scheme for both cases.

\subsection{Erasure model}
In an erasure channel with input $\Xk$ and output $\Xj$, the $n$-th component of $\Xj$ is either equal to the $n$-th component of $\Xk$, or erased with probability $p$.
The rate needed to decode $\Xk$ from $\Xj$  depends on the probability $p$ that an element is erased. We consider that the set $\Jsetk$ is made of $6$ sources $X_{(j_1)}, \ldots, X_{(j_6)}$, with respective erasure probabilities $p_{(j_1)} = 0.01$, $p_{(j_2)} = 0.05$,   $p_{(j_3)} = 0.1$, $p_{(j_4)} = 0.15$, $p_{(j_5)} = 0.2$ and $p_{(j_6)} = 0.25$. 
In our experiment, we consider $n=40000$. The source $\Xk$ is encoded with an LDPC staircase code \cite{Roca12RFC6816}. The systematic bits are discarded and only the parity bits are stored on the server. At the decoder, the parity bits are used to build $\Xk$ using the source $\Xjs$ in memory as side information, \emph{i.e.,} taking it for systematic bits and using the received parity bits to complete the codeword. Regarding the stored bitstream, the least correlated source $X_{(j_6)}$ corresponds to the highest bitrate and we store the amount of bits that leads to a very low error reconstruction of $\Xk$ from $X_{(j_6)}$. At the transmission stage, for each source $\Xj$ in memory, we extract from this bitstream the minimum amount of bits that are required to obtain $\Xk$ losslessly. The obtained rates are shown in Table \ref{table:erasure}. We see that the bitrate obtained with the LDPC staircase are close to the theoretical bounds. An important observation is also that from the same stored bitstream, a user is able to extract the piece of information that is required by its own navigation. In other words, the storage considers the worst case, but the rate is actually not penalized by the randomness of the navigation.

\subsection{Binary error model}
Consider a binary symmetric channel with input $\Xk$, output $\Xj$, and error probability $p = \P(\Xk \neq \Xj)$.
As before, we consider that the set $\Jsetk$ is made of $6$ sources $X_{(j_1)}, \ldots, X_{(j_6)}$, with respective error probabilities $p_{(j_1)} = 0.01$, $p_{(j_2)} = 0.05$,   $p_{(j_3)} = 0.1$, $p_{(j_4)} = 0.15$, $p_{(j_5)} = 0.2$ and $p_{(j_6)} = 0.25$. The compression rate depends on the error probability through the formula $H = -p\log_2 p-(1-p)\log_2(1-p)$.
For this channel, we encode the source $\Xk$ with an LDPCA code \cite{Varodayan06Eurasip}. As before, the parity bits of the least correlated source $X_{(j_6)}$ are stored and contain the parity bits of all the sources. 
From this bitstream, a part of the parity bits are extracted to transmit $\Xk$ given that one of the sources  $X_{(j_1)}, \ldots, X_{(j_6)}$ is available as side information at the decoder. The obtained rate are shown in Table~\ref{tab:bsc} for the respective number of symbols $n=396$ and $n=6336$.

As in the case of the erasure channel, the results of Table~\ref{tab:bsc} show that here again, the proposed implementation enables to extract a sub-bitstream that depends on the source $\Xj$ and to reach a very low reconstruction error probability of $\Xk$. 
This important result validates the idea that the rate is not penalized by the randomness of the user navigation.

%
\section{Conclusion}
In this paper, we have introduced and formalized the Sequential Massive Random Access problem, that deals with a set of sources encoded once for all on a server, followed by a partial transmission of the requested sources only. 
We have introduced a coding scheme in two parts that are the offline encoder for source storage, and the online extractor for transmission of the requested sources only.
From both theoretical and simulation results, we have shown that, for a reasonable storage cost, the partial transmission of the requested sources can be done with the same rate as if re-encoding was allowed during the online phase. 
This fundamental result may be transposed to all the practical SMRA scenarios, by deriving specific correlation models coupled with efficient channel codes. Some other interesting aspects of SMRA might be further developed too, such as the influence of the navigation graph on the storage and transmission rates.

\section*{Acknowledgement}
This work has received a French government support granted to the Cominlabs excellence laboratory and managed by the National Research Agency in the ``Investing for the Future'' program under reference ANR-10-LABX-07-01.




\bibliographystyle{IEEEtran}
\bibliography{abbr,biblio}

\begin{thebibliography}{10}
\providecommand{\url}[1]{#1}
\csname url@samestyle\endcsname
\providecommand{\newblock}{\relax}
\providecommand{\bibinfo}[2]{#2}
\providecommand{\BIBentrySTDinterwordspacing}{\spaceskip=0pt\relax}
\providecommand{\BIBentryALTinterwordstretchfactor}{4}
\providecommand{\BIBentryALTinterwordspacing}{\spaceskip=\fontdimen2\font plus
\BIBentryALTinterwordstretchfactor\fontdimen3\font minus
  \fontdimen4\font\relax}
\providecommand{\BIBforeignlanguage}[2]{{%
\expandafter\ifx\csname l@#1\endcsname\relax
\typeout{** WARNING: IEEEtran.bst: No hyphenation pattern has been}%
\typeout{** loaded for the language `#1'. Using the pattern for}%
\typeout{** the default language instead.}%
\else
\language=\csname l@#1\endcsname
\fi
#2}}
\providecommand{\BIBdecl}{\relax}
\BIBdecl

\bibitem{Hilbert_M_2011_science}
M.~Hilbert and P.~Lopez, ``The world's technological capacity to store,
  communicate, and compute information,'' \emph{Science}, vol. 332, no. 6025,
  pp. 60--65, Apr. 2011.

\bibitem{Tanimoto_2012_ieee-spm_ftv_fvt}
M.~Tanimoto, ``{FTV}: Free-viewpoint television,'' \emph{IEEE Signal Processing
  Magazine}, vol.~27, no.~6, pp. 555--570, Jul. 2012.

\bibitem{Roca12RFC6816}
\BIBentryALTinterwordspacing
V.~Roca, M.~Cunche, and J.~Lacan, \emph{{LDPC-Staircase Forward Error
  Correction (FEC) Schemes for FECFRAME}}, IETF Std. RFC 6816, Dec. 2012.
  [Online]. Available: \url{http://datatracker.ietf.org/doc/rfc6816/}
\BIBentrySTDinterwordspacing

\bibitem{Varodayan06Eurasip}
D.~Varodayan, A.~Aaron, and B.~Girod, ``Rate-adaptive codes for distributed
  source coding,'' \emph{EURASIP Signal Processing}, vol.~86, no.~11, pp.
  3123--3130, 2006.

\bibitem{maddah14IT}
M.~A. Maddah-Ali and U.~Niesen, ``Fundamental limits of caching,'' \emph{IEEE
  Transactions on Information Theory}, vol.~60, no.~5, pp. 2856--2867, 2014.

\bibitem{hassanzadeh16arxiv}
P.~Hassanzadeh, A.~Tulino, J.~Llorca, and E.~Erkip, ``Cache-aided coded
  multicast for correlated sources,'' \emph{arXiv preprint arXiv:1609.05831},
  2016.

\bibitem{dimakis10IT}
A.~G. Dimakis, P.~B. Godfrey, Y.~Wu, M.~J. Wainwright, and K.~Ramchandran,
  ``Network coding for distributed storage systems,'' \emph{IEEE Transactions
  on Information Theory}, vol.~56, no.~9, pp. 4539--4551, 2010.

\bibitem{Gamal82IT}
A.~{El Gamal} and T.~Cover, ``{Achievable rates for multiple descriptions},''
  \emph{IEEE transaction on Information Theory}, vol.~28, no.~6, pp. 851--557,
  Nov 1982.

\bibitem{diggavi04ITW}
S.~N. Diggavi and V.~A. Vaishampayan, ``On multiple description source coding
  with decoder side information,'' in \emph{IEEE Information Theory Workshop},
  2004, pp. 88--93.

\bibitem{sgarro77IT}
A.~Sgarro, ``{Source coding with side information at several decoders},''
  \emph{IEEE Transactions on Information Theory}, vol.~23, no.~2, pp. 179--182,
  1977.

\bibitem{Draper04}
S.~C. Draper, ``Universal incremental slepian-wolf coding,'' in \emph{{Allerton
  Conference on Communication, control and computing}}, 2004, pp. 1757 -- 1761.

\bibitem{Maugey_T_2013_tmm_int_mvsncnd}
T.~Maugey and P.~Frossard, ``Interactive multiview video system with low
  complexity 2d look around at decoder,'' \emph{IEEE Transactions on
  Multimedia}, vol.~15, pp. 1--13, Aug. 2013.

\bibitem{Slepian73IT}
D.~Slepian and J.~Wolf, ``{Noiseless coding of correlated information
  sources},'' \emph{IEEE Transactions on Information Theory}, vol.~19, no.~4,
  pp. 471--480, July 1973.

\bibitem{Wyner74IT}
A.~Wyner, ``{Recent Result in the Shannon theory},'' \emph{IEEE Transactions on
  Information Theory}, vol.~20, no.~1, pp. 2--10, 1974.

\bibitem{Draper07}
S.~C. Draper and E.~Martinian, ``{Compound conditional source coding,
  Slepian-Wolf list decoding, and applications to media coding},'' in \emph{{
  IEEE International Symposium on Information Theory}}, 2007.

\bibitem{Cheung_G_2011_tip_int_ssmvurfs}
G.~Cheung, A.~Ortega, and N.~Cheung, ``Interactive streaming of stored
  multiview video using redundant frame structures,'' \emph{IEEE Transactions
  on Image Processing}, vol.~3, no.~3, pp. 744--761, Mar. 2011.

\bibitem{yang10IT}
E.~Yang and D.~He, ``{Interactive encoding and decoding for one way learning:
  Near lossless recovery with side information at the decoder},'' \emph{{IEEE
  Transactions on Information Theory}}, vol.~{56}, no.~{4}, pp. {1808--1824},
  {2010}.

\bibitem{cover2006b}
T.~Cover and J.~Thomas, \emph{Elements of information theory, second
  Edition}.\hskip 1em plus 0.5em minus 0.4em\relax Wiley, 2006.

\bibitem{yeung2010}
R.~W. Yeung, \emph{{Information Theory and Network Coding}}.\hskip 1em plus
  0.5em minus 0.4em\relax Springer, 2010.

\bibitem{han2003b}
T.~Han, \emph{{Information-spectrum methods in information theory}}.\hskip 1em
  plus 0.5em minus 0.4em\relax {Springer}, {2003}.

\bibitem{MoulinIT07}
P.~Moulin and Y.~Wang, ``Capacity and random-coding exponents for channel
  coding with side information,'' \emph{IEEE Transactions on Information
  Theory}, vol.~53, no.~5, pp. 1326--1347, 2007.

\bibitem{dupraz:Tcom14}
E.~Dupraz, A.~Roumy, and M.~Kieffer, ``Source coding with side information at
  the decoder and uncertain knowledge of the correlation,'' \emph{IEEE
  Transactions on Communications}, vol.~62, no.~1, pp. 269 -- 279, Jan. 2014.

\bibitem{tian2007multistage}
C.~Tian and S.~N. Diggavi, ``On multistage successive refinement for wyner--ziv
  source coding with degraded side informations,'' \emph{IEEE Transactions on
  Information Theory}, vol.~53, no.~8, pp. 2946--2960, 2007.

\bibitem{watanabe14IT}
S.~Watanabe and S.~Kuzuoka, ``{Universal Wyner--Ziv coding for distortion
  constrained general side information},'' \emph{IEEE Transactions on
  Information Theory}, vol.~60, no.~12, pp. 7568--7583, 2014.

\bibitem{Feder02ITW}
M.~Feder and N.~Shulman, ``{Source broadcasting with unknown amount of receiver
  side information},'' in \emph{{Information Theory Workshop (ITW),
  Proceedings.}}, march 2002, pp. 302 -- 311.

\bibitem{Shulman03phd}
N.~Shulman, ``Communication over an unknown channel via common broadcasting,''
  Ph.D. dissertation, Tel Aviv University, 2003.

\bibitem{MacKay03ITbook}
D.~MacKay, \emph{Information Theory, Inference, and Learning Algorithms}.\hskip
  1em plus 0.5em minus 0.4em\relax Cambridge University Press, 2003.

\bibitem{MacKay05Fountain}
------, ``{Fountain Codes},'' \emph{IEE Proceedings on Communications}, vol.
  152, no.~6, pp. 1062--1068, Dec. 2005.

\bibitem{DragottiBook09DSC}
P.~L. Dragotti and M.~Gastpar, \emph{Distributed Source Coding: Theory,
  Algorithms and Applications}.\hskip 1em plus 0.5em minus 0.4em\relax Academic
  Press, 2009.

\end{thebibliography}

\end{document}